\documentstyle[12pt,epsfig]{article}
\def\beq{\begin{equation}}   \def\eeq{\end{equation}}
\def\bea{\begin{eqnarray}}   \def\eea{\end{eqnarray}}

\newcommand{\gsim}{\lower.7ex\hbox{$
\;\stackrel{\textstyle>}{\sim}\;$}}
\newcommand{\lsim}{\lower.7ex\hbox{$
\;\stackrel{\textstyle<}{\sim}\;$}}

\newcommand{\ra}{\rightarrow}
\newcommand{\mvec}[1]{|\vec{#1}\,|}

\newcommand{\Lam}{\Lambda_{\rm QCD}}

\renewcommand{\Im}{\mbox {Im}\:}

\newcommand{\bibit}[1]{\bibitem{#1}}

\newcommand{\GeV}{\,\mbox{GeV}}

\newcommand{\matel}[3]{\langle #1|#2|#3\rangle}
\newcommand{\aver}[1]{\langle #1\rangle}

\renewcommand{\sp}{ {\rm sp} }

\begin{document}

\def\lsim{\mathrel{\rlap{\lower3pt\hbox{\hskip0pt$\sim$}}
    \raise1pt\hbox{$<$}}}         
\def\gsim{\mathrel{\rlap{\lower4pt\hbox{\hskip1pt$\sim$}}
    \raise1pt\hbox{$>$}}}         

\begin{titlepage}
\renewcommand{\thefootnote}{\fnsymbol{footnote}}

\begin{flushright}
UND-HEP-99-BIG\hspace*{.2em}02\\
hep-ph/9902315\\
\end{flushright}
\vspace{.3cm}

\begin{center} \Large
{\bf
Heavy Quark Expansion and \\ Preasymptotic Corrections to
Decay Widths\\ in the 't Hooft Model
}
\end{center}
\vspace*{.3cm}
\begin{center}
{\Large
Ikaros Bigi$^{\:a}$ and Nikolai Uraltsev$^{\:a,b}$
\\
\vspace{.4cm}
{\normalsize
$^a${\it Dept.\ of Physics,
Univ.\ of Notre Dame du
Lac, Notre Dame, IN 46556, U.S.A.}\\
$^b${\it Petersburg Nuclear Physics Institute,
Gatchina, St.\,Petersburg, 188350, Russia
}
}
}

\vspace*{1.7cm}
{\large \bf Abstract}
\vspace*{.25cm}

\end{center}

\thispagestyle{empty}
\setcounter{page}{0}


\noindent
We address nonperturbative power corrections to  inclusive decay widths
of heavy flavor hadrons in the context of the 't~Hooft  model
(two-dimensional QCD at $N_c \to \infty$), with the emphasis on the
`spectator-dependent' effects, {\it i.e.} those sensitive to the flavor of
the spectator.
The summation of exclusive widths  is performed analytically using
the 't~Hooft  equation. We show that the $1/m_Q$ expansion of both the
Weak Annihilation  and Pauli Interference widths coincides with the OPE
predictions, to the computed orders. Violation of local duality in the
inclusive widths is quantified, and the new example is identified where
the OPE prediction  and the actual effect are completely saturated by a
single final state. The qualitative aspects of quark hadronization
emerging from the analysis in the 't~Hooft model are discussed.

\noindent
Certain aspects of summation of spectator-independent hadronic weak
decay widths are given in more detail, which were not spelled out
previously.
We also give some useful details of the $1/m_Q$ expansion in the
't~Hooft model.
\vspace*{.4cm}

\noindent
PACS numbers: 12.38.Aw, 12.39.Hg, 23.70.+j, 13.35.Dx

\vspace*{.2cm}
\vfill
\end{titlepage}

\newpage

\tableofcontents

\section{Introduction}

The decays of heavy flavor hadrons $H_Q$ are shaped by nonperturbative
strong interaction dynamics which, at first sight, completely obscures
most of the properties of the underlying weak interactions self-manifest
at the
quark level. It is suffice to say that the actual hadrons, rather than
quarks are observed in the final state.
The actual
dynamics of confinement in QCD to a large extent remains mysterious.
Nevertheless, significant progress has been achieved in describing
heavy flavor decays applying the formalism based on Wilson's operator
product expansion (OPE) \cite{wilson}. In particular, it became possible
to quantify the effects of the confining domain on the inclusive decay
rates. This theory is in the mature stage now (see Refs.~\cite{stone2,rev}
and references therein).

Among the general statements derived for the heavy quark decays, we
mention here

$\bullet$ Absence of $\Lam /m_Q$ corrections to all types of fully
inclusive decay widths \cite{buv,bs}.\footnote{The OPE for the inclusive
widths, actually, is {\it a priori} governed by the energy release
rather than literally $m_Q$ \cite{volshif}. For simplicity, we do not
distinguish between them parametrically unless it becomes essential.}

$\bullet$ The leading nonperturbative corrections arise in order
$1/m_Q^2$ and are given by the expectation values $\mu_\pi^2,\; \mu_G^2$
of the two heavy quark operators, kinetic and chromomagnetic.
While the first effect
is universal amounting to the correction $-\mu_\pi^2/2m_Q^2$,
the Wilson coefficient for the second one depends on the considered
process. Both, however, are insensitive to the flavor of the
spectator(s) (``flavor-independent'' corrections) \cite{buv,bs}.

$\bullet$ The widths are determined by the short-distance running quark
masses $m_Q(\mu)$ \cite{pole}. These are shielded against uncontrollable
corrections from the infrared domain which would otherwise bring in
uncertainty $\delta m_Q/m_Q \sim \Lam/m_Q$.

$\bullet$ The effects sensing the spectator flavor {\it per se}, emerge
at the level $1/m_Q^3$ \cite{vsold,mirage,buv}. They are conventionally
called Weak Annihilation
(WA) in mesons, Weak Scattering (WS) in baryons and Pauli Interference
(PI) in both systems. Their magnitudes are given by the expectation values of
local four-quark operators.\footnote{In the context of the heavy quark
expansion, local operators have a more narrow meaning denoting the
generic operator of the form $\bar{Q}O Q$, with $O$ being a local operator
involving only light degrees of freedom.}

For practical applications we should keep the following in mind
(for a recent dedicated discussion, see Refs.~\cite{D2,inst,rev}):

-- Good control over the perturbative expansion must be established to
address power-suppressed effects.

-- The consistent Wilsonian OPE requires introducing the
separation of ``hard'' and ``soft'' scales, with the borderline $\mu$
serving as the normalization point in the effective theory.

-- One has to allow, in principle, for short-distance (small-coupling
regime) effects that are not directly expandable in the powers of the
strong coupling.

-- Account must be taken of the fact that the OPE power series are only
asymptotic \cite{shifcont}, and reconstructing
from them the actual Minkowskian
observable, generally speaking, potentially leaves out the oscillating
(sign-alternating) contributions suppressed, in  a certain interval of
energies, by only a power of the high momentum scale.
This is compounded by the fact that in practice one can typically
determine only the first few terms in the power expansion.

The last item in the list is behind the phenomenon of violation
of local parton-hadron duality; in many cases it is among the primary factors
potentially limiting the accuracy of the theoretical expansion.

In the actual QCD these technical complications are often interrelated.
Therefore, it is instructive to investigate the OPE in a simplified
setting where these elements can be disentangled. As explained in
Ref.~\cite{D2}, this is achieved in
QCD formulated in
1+1 dimensions. Additionally, employing the limit $N_c \to \infty$ one
arrives at the exactly solvable 't~Hooft model where all the features
can be traced explicitly. It is important that the 't~Hooft model
maintains the crucial feature of QCD -- quark confinement -- which
is often believed to be tightly related to the violation of local
duality. Yet in 1+1 dimensions confinement appears already in the
perturbative expansion.

The 't~Hooft model has often been used as a theoretical laboratory for
exploring various field-theoretic approaches \cite{exploit}. Most recently
the OPE for the inclusive
widths and the related sum rules in the heavy flavor transitions
\cite{optical} were analytically studied in Ref.\,\cite{D2},
where a perfect
match between the OPE power expansion and the actual asymptotics of the
widths was found. The known high-energy asymptotics of the
spectrum in the model allowed us to determine the
violation of local duality in the inclusive widths at large $m_Q$. As
expected, it obeyed the general constraints imposed by the OPE.
Moreover, at least in the framework of this simplified model, the main
features of duality violation could be inferred
from the
parton-level analysis itself, the working tools of the OPE. The
suppression of the
duality-violating component in $\Gamma_{H_Q}$ was
found to be rather strong, with the power of
$1/m_Q$, however, depending
essentially on the particulars of the considered model and the process.

Ref.\,\cite{D2} focussed on
flavor-independent effects.
To this end it was assumed that the spectator quark $q_{\rm sp}$ has a
flavor different from all quarks in the final state,
thus ruling out both WA
and PI. The OPE analysis of these effects is also
straightforward. Nevertheless, they may be of independent
interest for several reasons.

First, WA and PI represent power-suppressed and thus purely
preasymptotic effect. In such a situation
one may expect a later onset of duality and
more significant violations of local duality. Since the above effects
are numerically enhanced for actual charm and beauty hadrons, studying
this question has practical importance.

\thispagestyle{plain}
\begin{figure}[hhh]
 \begin{center}
 \mbox{\epsfig{file=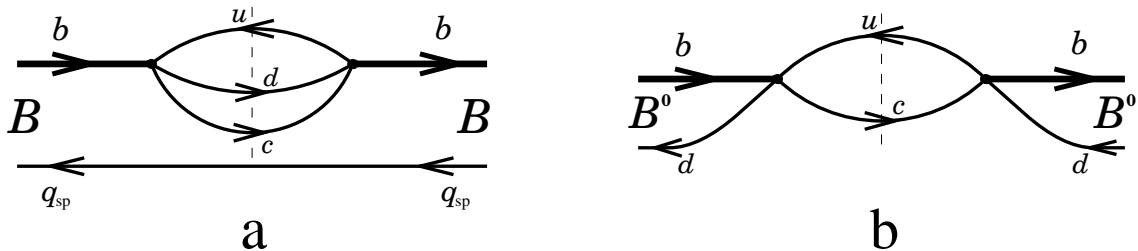,width=15cm}}
 \end{center}
 \caption{
a) Quark diagram describing the leading quasifree term in the decay
width.
b) Bare quark diagram for WA.
}
\end{figure}

Another reason to look more closely at the spectator-dependent effects
is related to the color-flow considerations usually employed in the
context of the large-$N_c$ perspective on QCD, and interpreting the OPE
predictions in terms of hadronic states. In the case of the
quasi-free quark decay width or the WA processes one finds a rather
straightforward correspondence between the OPE expressions and the
hadronic contributions already in the simplest quark picture where quark
allocation over the final state hadrons is unambiguous
(such a description is expected to hold at $N_c \to \infty$). Let us
consider, for example,
the free parton decay diagram Fig.\,1a. The $\bar{u}d$
pair is in a colorless state and typically has a large momentum
$q^2 \sim m_b^2$ flowing through it. It is then naturally dual to the
contributions from the hadronic resonances in the $V$--$A$ channel
(in particular when integrated over $q^2$), much in the same way as
in ${\rm e}^+{\rm e}^-$
annihilation or hadronic $\tau$ decays. The $c$ quark together with
the spectator antiquark produces another string of hadronic excitation.
Furthermore, the interaction between these
two hadronic clusters can naturally be small at large $m_b$. WA,
Fig.\,1b, looks even simpler in this respect; we will discuss it in
detail later on.

\thispagestyle{plain}
\begin{figure}[hhh]
 \begin{center}
 \mbox{\epsfig{file=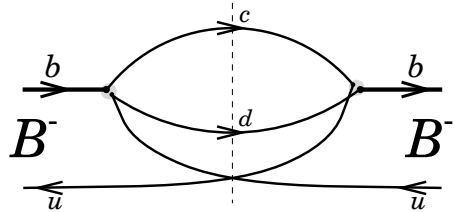,width=6cm}}
 \end{center}
 \caption{
Quark diagram for PI in $B$ meson decays. The weak vertices are broken
to show the color flow yielding the leading-$N_c$ contribution.
}
\end{figure}

The hadronic picture of the processes underlying PI {\it a priori}\,
is less obvious, Fig.\,2. The $\bar{u}$ quark produced in the decay must be
slow to interfere with the valence $\bar{u}$. The large momentum here flows
through the diquark loop ($cd$) which therefore represents the ``hard core''
of the process. The practical OPE, effectively, prescribes to replace
the propagation of this diquark by a nearly free di-fermion loop,
which amounts to evaluating its absorptive part as if
the production of the free quarks was considered. Basically, no distinction
emerges
compared to the color-singlet $\bar{q}q'$ pairs in Figs.\,1. This
may leave one with the feeling of discomfort, for no colored states (in
particular, with the diquark content) is present in the physical
spectrum. In other words, the diquark configuration {\it per se} cannot
be dual to the mesonic states at any arbitrary large momentum transfer.

Alternatively, one can combine a ``hard'' quark from the loop in
Fig.\,2 with the slow spectator antiquark to have a color-singlet meson-like
configuration. However, such a pair naively is not ``hard'':  at least
in the perturbative partonic picture with $p_{\rm sp}\sim m_{\rm sp} \to
0$ its invariant mass vanishes irrespective of $m_b$.
While such reasoning is clearly of the hand-waving variety, it
illustrates nevertheless that interference effects are more subtle.

A more troublesome feature of the interference is also illustrated by the
observation made in the early 90s by Shifman \cite{shifman}. He
considered a more general
scenario with
both charged- and neutral-current type interactions, as described by the
effective weak Lagrangian
\beq
{\cal L} = -\frac{G}{\sqrt{2}} \left[a_1\,(\bar c \gamma_\mu
(1\!-\!\gamma_5)b)
\,
(\bar{d}\gamma^\mu (1\!-\!\gamma_5)u ) \,+\,
a_2\,(\bar d \gamma_\mu (1\!-\!\gamma_5)b) \,
(\bar{c}\gamma^\mu (1\!-\!\gamma_5)u )\right]
\,+\, {\rm H.c.}
\;.
\label{5}
\end{equation}
The leading (rather than the power-suppressed spectator-dependent) width
was addressed.
The parton result depends on the color factors $a_1$, $a_2$ in the
following way:
\beq
\Gamma_Q\;\sim\; N_c\left( a_1^2 +a_2^2+\frac{2}{N_c} a_1 a_2\right)
\;.
\label{6}
\end{equation}
On the other hand, the usual counting rules yield the decay amplitudes
into the two-meson final states in the form
\bea
\nonumber
{\cal M} \propto \sqrt{N_c} \left( a_1 +\frac{1}{N_c}a_2\right) &
\mbox{ for} & \mbox{ ``$D_s \pi^-$'' states}
\\
{\cal M} \propto \sqrt{N_c} \left( a_2 +\frac{1}{N_c}a_1\right) &
\mbox{ for}  & \mbox{ ``$D K$'' states},
\label{7}
\eea
where, for illustrative purposes, we call $q_{\rm sp}$ the strange quark
to simplify distinguishing between
the two different ways to pair the quarks into mesons.
(Since we discuss the leading free-parton amplitude,
the flavor of the spectator is chosen to be different from all other
quarks in the process.)
Adopting the rules Eq.\,(\ref{7}) one gets
\beq
\Gamma_Q\;\sim\; N_c\left( a_1^2 +a_2^2+\frac{4}{N_c} a_1 a_2\right)
\label{8}
\end{equation}
more or less independently of the dynamics. While the dependence for the
terms $\sim a_1^2$ and $\sim a_2^2$ is reproduced, there is a clear
mismatch between Eqs.\,(\ref{6}) and (\ref{8}) in the term
describing the interference of the two different color amplitudes
\cite{shifman}.

There is little doubt that the formal OPE asymptotics must work at arbitrary
$N_c$.
The arguments above might suggest, however, that the onset of duality
is delayed for suppressed effects, for example, grow with $N_c$.

In reality, we do not think that there is convincing evidence
supporting such reservations about applying the OPE to flavor-dependent
corrections. To provide an additional justification,
we have explicitly analyzed both PI and WA in the 't~Hooft model. We have found
complete consistency with the OPE, with the onset of duality largely
independent of the details. As a matter of fact, the parton-deduced OPE
expression for PI appears to be {\em exact} in the chiral limit when all
involved quarks (but $Q$) are massless. The resolution of the above
paradoxes emerges in a rather straightforward manner as well; we will comment
on them in subsequent sections.

We note that we disagree with the claims of the recent paper \cite{gl2}
which found a mismatch between the actual WA width and the OPE-based
prediction, relying on numerical computations. We have determined the
leading effect analytically and showed it to coincide with the OPE
result. We comment on the apparent drawbacks in the analysis of
Ref.\,\cite{gl2} in Sect.\,6.

The paper is organized as follows. After this introduction, in Sect.\,2 we
sketch the aspects of the 't~Hooft model important for
addressing weak decays.
In Sect.\,3. we analyze the
effects of WA at $N_c\to \infty$ and analytically compute the
large-$m_Q$ asymptotics of the corresponding
heavy meson weak decay width, with
technical details
given in Appendix~1.
Sect.\,4 addresses PI; we analytically compute this width up to terms
like $\Gamma_{\!H_Q}/m_Q^3$ and find full agreement with the
expressions obtained in the OPE. The effects of local duality violation
at large $m_Q$ are quantified. The special case -- with massless
final-state quarks -- is identified where duality violation is totally
absent from the spectator-dependent part of the width. In Sect.\,5 we
present a more detailed derivation of the total decay width up to
corrections $1/m_Q^3$ explicitly accounting for nonzero light-quark
masses, to demonstrate consistency with the OPE (a detailed description of
this
analysis had been omitted from Ref.\,\cite{D2}). Sect.\,6
comments on the
analyses which have claimed observing inapplicability of the OPE
predictions based on numerical computations. Sect.\,7 comprises
conclusions and overlook and outlines our perspective on the problem of OPE
and duality violation in the decays of heavy flavor hadrons.

Most technicalities are relegated to Appendices.
Appendix~2 collects a number of relations useful
in constructing analytic $1/m_Q$ expansion in the 't~Hooft model and
summing the exclusive widths. In particular, we give simple
expressions for the leading terms in the transition amplitudes in
Appendix~2.2, perform the differential fixed-$q^2$ semileptonic decay width
summation up to $1/m_Q^2$ corrections in Appendix~2.3, prove the OPE
prescription for the domain of large $q^2$ and demonstrate the proper
functional form of the transition amplitudes in Appendix~2.4. The
expression for the IW functions in terms of the 't~Hooft eigenfunctions
is quoted in Appendix~3. Appendix~4 reports a direct covariant
computation of the
perturbative radiative corrections performed while working on paper
\cite{D2}; it shows that the result coincides with what is obtained by
summing exclusive decay channels.

\section{The 't Hooft model and heavy quark decays}

The 't Hooft model, the 1+1 QCD with $N_c \to \infty$ has been described in
many papers \cite{thooft,callan,einhorn,ff}.
The first dedicated studies of heavy
quarks in the 't~Hooft model date back to the early 90s \cite{burkswan,burk}.
Recent paper \cite{D2} specifically addressed heavy quark decays and the
OPE in this model. Here we only recapitulate some basic features.

The Lagrangian has the form
\begin{equation}
{\cal L}_{1+1}=-\frac{1}{4g_s^2} \,
G_{\mu\nu}^a G_{\mu\nu}^a \,+\, \sum
\bar\psi_i
(i\not \!\!D -m_i)\psi_i \; , \; \;\;\;
i D_\mu=i\partial_\mu + A_\mu^a T^a\, .
\label{9}
\end{equation}
The coupling $g_s$ has dimension of mass. With the above normalization
of the gauge field, $A_\mu$ still has dimension of mass as in $D=4$. The
fermion fields $\psi(x)$, however, carry dimension of $m^{1/2}$.

The OPE analysis is carried out universally for arbitrary number of
colors, and so far $N_c$ is kept finite. Anticipating the large $N_c$
limit for the final analysis, we define a parameter $\beta$
\begin{equation}
\beta^2\;=\;\frac{g_s^2}{2\pi}\,\left( N_c -\frac{1}{N_c} \right)\;,
\label{10}
\end{equation}
that remains finite at $N_c \to \infty$. It plays the role of the
nonperturbative scale $\Lam$.

Following the actual Standard Model, we choose the weak decay interaction of
the current-current  form. Since in $D=2$ the axial current is
related to the vector one, $J_\mu^A= \epsilon_{\mu\nu} J_\nu^V$, we
simply consider the $V$$\times$$V$ interaction:
\begin{equation}
{\cal L}_{\rm weak} \;=\;-\frac{G}{\sqrt{2}}\,(\bar q \gamma_\mu Q)
\,
(\bar{\psi}_a\gamma^\mu \psi_b )\; + \; {\rm H.c.}\;,
\label{11}
\end{equation}
where the dimensionless $G$ is an analogue of the Fermi constant. For
semileptonic decays $\psi_{a,b}$ are colorless (leptonic) fields. In
what follows our main interest lies in nonleptonic decays with
$\psi_{a,b}$ being the quark fields. To make the notations more
transparent, we adhere to the cases of interest in actual QCD and
denote the $\psi$ fields as $u$ and $d$ quarks, while $Q$ will be a
synonym of the $b$ quark, and $q$ called $c$ quark (whether we chose
$m_q \gg \Lam$ or consider $m_q \lsim \Lam$). The spectator quark
$q_{\rm sp}$ can be either $u$ or $d$ (for studying WA or PI), or
different in flavor from both.

To address inclusive widths of a heavy flavor hadron $H_Q$ one considers the
forward transition
amplitude appearing in the second order in the decay interaction
\cite{vsold}:
\begin{equation}
\Gamma_{H_Q}\;=\;2\,{\rm Im} \,\int {\rm d}^D x \,
\frac{1}{2M_{H_Q}} \langle H_Q|\,i\,T\left\{ {\cal L}_{\rm weak}(x)
{\cal L}_{\rm weak}^\dagger(0)\right\}|H_Q\rangle
\;.
\label{12}
\end{equation}
In the limit $N_c\to\infty$, with $H_Q$ being the mesonic $(Q\bar q_{\rm
sp})$ states, factorization of the amplitudes holds, which takes the
following form for the transition operator:
\begin{equation}
\int {\rm d}^D x \,
\langle H_Q|\,i\,T\left\{ {\cal L}_{\rm weak}(x)
{\cal L}_{\rm weak}^\dagger(0)\right\}|H_Q\rangle
\;=\;
\frac{G^2}{2}\, \int {\rm d}^D x \, T^{\mu\nu}(x)\, \Pi_{\mu\nu}(x)
\;,
\label{14}
\end{equation}
where we have introduced the ``semileptonic'' $T_{\mu\nu}$ and ``hadronic''
$\Pi_{\mu\nu}$ tensors:
\bea
\label{15}
\Pi_{\mu\nu}(x)\; &=& \;\langle 0|i\,T \left \{ \bar{d}(x) \gamma_\mu
u(x) \, \bar{u}(0) \gamma_\nu d(0)\right\}|0\rangle\;,
\\
\label{16}
T^{\mu\nu}(x)\;&=& \; \langle H_Q |i\, T \left \{\bar{q}(x) \gamma^\mu Q(x)\,
\bar{Q}(0) \gamma^\nu q(0)\right\}|H_Q\rangle\;.
\eea
The Cutkosky rules then yield
\begin{equation}
\Gamma_{H_Q}\;=\;
G^2\, \frac{1}{M_{H_Q}} \, \int {\rm d}^D x \, {\rm Im}\,T^{\mu\nu}(x)\,
{\rm Im}\, \Pi_{\mu\nu}(x)
\;.
\label{17}
\end{equation}

The factorized representation of the decay width holds only at
$N_c\to\infty$ where the momenta of the $\psi_a\bar\psi_b$-pair and
$(q\bar q_{\rm sp})$ become observables separately. In other words, in
this limit there is a rigid quark allocation over
the particular hadronic final state and factorization of the corresponding
amplitudes, and there is no  ``cross-talk'' between them.  Yet,
Eq.\,(\ref{17}) represents a certain observable at arbitrary $N_c$ and,
as such, enjoys the full rights of being studied regardless of the
details of the model. In particular, at large energy release it is a
short-distance observable and can be subjected to an OPE anatomy. In
what follows we will discuss this quantity
and refer to it as the inclusive decay width as motivated by the
large-$N_c$ limit.

It is worth noting at this point that the qualitative difference between
nonleptonic and semileptonic inclusive widths disappears for
$N_c\to\infty$. The nonleptonic width is given directly in terms of the
differential semileptonic distributions
$\frac{{\rm d}\Gamma^{\rm sl}}{{\rm d} q^2}$ (though, in $D=2$ one may
have to consider the decays with massive leptons as well). Indeed, with
$m_u=m_d$ as an example, one has (in the momentum representation)
\begin{equation}
\Pi_{\mu\nu}(q^2) \;=\;
\frac{1}{\pi}\: \Pi(q^2)\,
\left(q^2\delta_{\mu\nu}-q_\mu q_\nu \right)\;,
\qquad
\rho(q^2)\;\equiv \;
-\frac{1}{\pi}\:\Im \Pi(q^2)\;,
\label{18}
\end{equation}
and
\begin{equation}
\Gamma_{H_Q}^{\rm nl}\;=\;
\int\, {\rm d}q^2 \: \rho(q^2)\,\Gamma_{\rm sl}(q^2)
\qquad \mbox{with }\;\,
\Gamma_{\rm sl}(q^2)\;=\;
\frac{1}{\rho_{\rm lept}(q^2)}\,
\frac{{\rm d}\Gamma_{H_Q}^{\rm sl}}{{\rm d} q^2}
\;.
\label{20}
\end{equation}

In $D=2$ the correlator of vector currents for {\em massless} quarks is
known exactly and is very simple:
\begin{equation}
\Pi(q^2) \;=\;
\frac{N_c}{q^2}
\;,
\qquad
\rho(q^2)\; = \; N_c \,\delta(q^2)\;.
\label{22}
\end{equation}
With nonzero quark masses the spectral density shifts upward, to the
mass scale $\sim \beta m$ or $m^2$. A high-energy tail in $\rho$
also appears $\sim N_c(m_u^2+m_d^2)/q^4$. This will be quantified in
Sect.\,3.

More specific for heavy quark decays is the ``semileptonic'' part
$T_{\mu\nu}(x)$, Eq.\,(\ref{16}). The general color counting rules
determine its $N_c$ behavior:
\begin{equation}
T_{\mu\nu}(x)\;\sim\; N_c^1\;.
\label{23}
\end{equation}
Such a leading contribution, however, can arise only with the vacuum as
intermediate state; all other contributions scale as $N_c^0$, or even
are further suppressed. The vacuum intermediate state is
possible only when the decay quark $q$ has the same flavor as
$q_{\rm sp}$. This is the effect belonging to WA. Therefore, one has
\bea
\label{25}
\Gamma_{\rm nl}^{\rm WA} \; & \sim \; N_c^2\;,\qquad \;
\Gamma_{\rm sl}^{\rm WA} \; & \sim \; N_c^1\\
\Gamma_{\rm nl} \;& \sim \; N_c^1\;,\qquad \;
\Gamma_{\rm sl} \; & \sim \; N_c^0 \qquad \mbox{ at } q\ne q_{\rm sp}\;.
\label{26}
\eea
Since WA is a leading-$N_c$ effect, vacuum factorization saturates
$T_{\mu\nu}^{\rm WA}$ at $N_c\to\infty$, and the effect takes the
simplest
form. This is the subject of the next section.

On the other hand, the ``usual'' non-spectator widths are formally
subleading in $N_c$ (even though they may yield the dominant
contribution to the decay width for a particular type of the heavy
meson). For such amplitudes the naive factorization does not hold, and
the explicit expressions take a far less trivial form. In the context of
the OPE, this emerges as ``color-disfavored'' structure of the resulting
local operators, so that {\it a priori} the factorization cannot be applied
to evaluate their expectation values \cite{D2}.

In the limit $N_c\to\infty$ the spectrum of $1$+$1$ QCD consists of
mesonic quark-antiquark bound states which are stable under
strong interactions. The meson masses are given by eigenvalues of the
't~Hooft equation
\begin{equation}
M^2_{n}\varphi_n(x) =
\left[
\frac{m_1^2 - \beta ^2}{x} + \frac{m_2^2 - \beta ^2}{1-x}
\right]
\varphi_n(x) - \beta ^2 \int_0^1{\rm d}y \,\frac{\varphi_n(y)}{(y - x)^2}\;,
\label{30}
\end{equation}
where $m_{1,2}$ are the bare quark masses of the constituents, and the
integral is understood in the principal value prescription.
The
solutions to the equation are the light-cone wave functions $\varphi
(x)$, with $x \,\in \,[0,1]$ having the meaning of the portion of
momentum carried by the (first) quark. They are singular at $x=0$ and
$x=1$ where their
behavior is given by $x^{\gamma_0}$ and $(1-x)^{\gamma_1}$,
respectively, with $\gamma_{0,1}$ defined by the following conditions:
\begin{equation}
\frac{\pi \gamma_0}{{\rm tan}\pi \gamma_0} =
-\frac{m_1^2-\beta^2}{\beta^2} \,\,, \;\;\;\;
\frac{\pi \gamma_1}{{\rm tan}\pi \gamma_1}=
-\frac{m_2^2-\beta^2}{\beta^2} \,.
\label{32}
\end{equation}
In full analogy with nonrelativistic quantum mechanics, the
eigenfunctions  $\varphi_n$  form a basis (complete in the physical
space):
\begin{equation}
\int_0^1 {\rm d}x\: \varphi_n(x)\varphi_k(x)\;=\; \delta_{nk}\;,
\qquad\;
\sum_n \,\varphi _n(x) \varphi _n(y) \;=\; \delta (x-y)\;.
\label{33}
\end{equation}

The weak decay constant of a particular meson is given by
\begin{equation}
f_n\;=\; \sqrt{\frac{N_c}{\pi}}\,
\int_0^1 {\rm d}x \:\varphi_n(x)\;,
\label{34}
\end{equation}
and the polarization tensor of vector currents (at $m_u=m_d$) takes the
form
\begin{equation}
\Pi(q^2)\;=\; \pi \sum_n \,\frac{f_n^2}{q^2-M_n^2}\;, \qquad
\rho(q^2)\;=\; \pi \sum_n \,f_n^2\,\delta\left(q^2-M_n^2\right)\;.
\label{35}
\end{equation}
As mentioned above, at $m_u=m_d=0$ one has $M_0=0$ and $f_0=
\sqrt{N_c/\pi}$, but for all excitations $f_n=0$.

The transition formfactors between two mesonic states that define the
non-annihi\-lation widths for large $N_c$, are of order $N_c^0$. Since the
weak quark currents $\bar{Q}q$ are formally of order $N_c^1$,
these formfactors are ``subleading'' in the same sense
as was discussed
previously and, in general have a more complicated form corresponding to
the first order correction in the $1/N_c$ expansion \cite{einhorn,ff}.

\section{Weak Annihilation at $N_c \to \infty$}

WA in the decays of heavy mesons becomes possible when one of the quarks
produced in the weak vertex has the same flavor as the spectator
antiquark. We assume $q=q_{\rm sp}$, in our notations. As detailed in
the preceding section, in this case there is a
single contribution to the transition tensor
$T_{\mu\nu}$ proportional to $N_c$ and leading to $\Gamma_{H_Q} \sim
N_c^2$. This is associated with the vacuum intermediate state, and is
given by \footnote{We neglect the contribution of another, two-particle
state $|H_Q(P) H_Q(P) \rangle$, also corresponding to vacuum
factorization,
but yielding the $u$-channel discontinuity.}
\beq
T_{\mu\nu}(x)\;=\;
i\, {\rm e\,}^{-iPx} \vartheta(x_0)
\:
\langle H_Q|\,\bar{Q}\gamma_\nu q\,|0\rangle
\langle 0|\,\bar{q}\gamma_\mu Q\,|H_Q\rangle
\;,
\label{47}
\end{equation}
(with $P_\alpha$ denoting the momentum of the decaying heavy flavor hadron
$H_Q$), so that, in the momentum representation,
\beq
{\rm Im}\,T_{\mu\nu}(q)\;=\; \frac{1}{2}
(2\pi)^D \,\delta^D(P-q)\:
\langle H_Q|\,\bar{Q}\gamma_\nu q\,|0\rangle
\langle 0|\,\bar{q}\gamma_\mu Q\,|H_Q\rangle
\;.
\label{48}
\end{equation}
This expression is valid in arbitrary dimension for any choice of the weak
current -- in general, one only must replace $\bar{Q}\gamma_\mu q$ by
the appropriate quark bilinear. Therefore, at $N_c \to \infty$ one has
\begin{equation}
\Gamma^{\rm WA}_{H_Q}\;=\; G^2 \:\frac{1}{2M_{H_Q}}\,
\langle H_Q|\,\bar{Q}\gamma_\mu q\,|0\rangle
\langle 0|\,\bar{q}\gamma_\nu Q\,|H_Q\rangle\:
{\rm Im}\,\Pi_{\mu\nu} \! \left(M_{H_Q}^2\right)
\;,
\label{56}
\end{equation}
which is illustrated in Fig.\,3.
It can be traced that the OPE corresponds to the same expression
if the expectation values of all the
higher-dimension four-quark operators reduce to their vacuum
factorized values (for earlier discussion of WA in a similar context,
see, Ref.\,\cite{WA}). The latter formally holds, in turn, at $N_c \to
\infty$.

\thispagestyle{plain}
\begin{figure}[hhh]
 \begin{center}
 \mbox{\epsfig{file=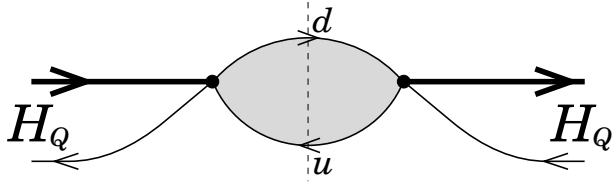,width=8cm}}
 \end{center}
 \caption{
WA correction to the inclusive decay width in the large-$N_c$ limit.
Shaded loop depicts the exact polarization operator.
}
\end{figure}

In $D=2$ for pseudoscalar $H_Q$ one has $\langle 0|\,\bar{q}\gamma_\mu
Q\,|H_Q\rangle = i f_{H_Q} \epsilon_{\mu\nu} P_\nu$. For simplicity, we
will further limit ourselves by the case $m_u=m_d$. Then
\begin{equation}
\Gamma^{\rm WA}_{H_Q}\;=\; \frac{G^2}{2} \: f_{H_Q}^2 M_{H_Q}^3\,
\rho(M_{H_Q}^2)\;.
\label{54}
\end{equation}

Strictly speaking, in practical applications of the OPE,
$\Pi_{\mu\nu}$ itself is usually likewise expanded in $1/m_Q^2$. Also, the
deviation of $M_{H_Q}^2$ from $m_Q^2$ and/or the values of $f_{H_Q}$ are
expanded around their asymptotic values at $m_Q \to \infty$. Therefore,
the sensible check of duality for practical OPE in WA in the
framework of the large-$N_c$ approximation is only comparison of the
actual behavior of ${\rm Im}\,\Pi(q^2)$ at large $q^2$ with its OPE
expansion obtained from the deep Euclidean domain.

For massless $u$ and $d$ quarks, the exact polarization operator of the
vector currents is given by Eq.\,(\ref{22}); the WA width,
therefore, vanishes. A non-zero result is obtained if one considers a
scalar (pseudoscalar) polarization operator, or if $m_u$ or $m_d$ do not
vanish. The absorptive part $\rho(q^2)$ is saturated by the comb
of narrow resonances with heights $\sim N_c$ and widths $\sim
1/N_c$. Therefore, the formal limit $N_c \to \infty$ requires an
alternative to point-to-point comparison of the actual hadronic
probabilities with the parton-calculated, or OPE-improved short-distance
expansion, even at arbitrary large energies. This implies a certain
smearing procedure for the actual hadronic probabilities.

Note that, according to Eq.\,(\ref{56}) the width -- however singular
it is -- always remains integrable around the resonances (see also the
discussion below, Eqs.\,(\ref{70}), (\ref{72}-\ref{77})). By virtue of
the dispersion relations the integral of the decay width is expressed
via the transition amplitude in the complex plane. This amplitude is
regular even in the formal limit $N_c \to \infty$ when the resonances
become infinitely narrow.

Smearing enters naturally when one considers the `imaginary' part
$\frac{1}{2i}\!\left[\Pi(s)\!-\!\Pi(s^*)\right]$
at complex $s$, somewhat away
from the physical cut at $s>0$. According to a dispersion relation it
amounts to averaging the physical cross section $R(s)$ with a specific
weight,
\begin{equation}
\frac{1}{2i}\left[\Pi(s)-\Pi(s^*)\right]\;=\; \frac{1}{\pi}\,
\int \,{\rm d}\tilde s \:\frac{\Delta}{(\tilde s-s_0)^2+\Delta^2}
\:R(\tilde s)\;,
\qquad \; s=s_0+i\Delta\,.
\label{70}
\end{equation}
One can also use different choices of the smearing function having
singularities away from the physical cut.

A similar procedure, in principle, is required for the inclusive decays
of heavy flavors. Strictly speaking, one must introduce the complex
variable $\omega$ to study the analytic properties of the transition
amplitude in question \cite{WA,inst,D2}:
\begin{equation}
{\cal A}(\omega)\;=\; \int\, d^Dx\; {\rm e}\,^{-i\omega(vx)}\:
\langle H_Q|\,i\,T\left\{ {\cal L}_{\rm weak}(x)
{\cal L}_{\rm weak}^\dagger(0)\right\}|H_Q\rangle
\;.
\label{72}
\end{equation}
It can be visualized as the transition amplitude governing the total
(weak) cross section of the scattering of a fictitious spurion particle
$S$ on the heavy quark,
\begin{equation}
S(q) + H_Q (p)\; \to \mbox{ light hadrons } \;,
\label{73}
\end{equation}
or the weak decay width in the process
\begin{equation}
Q\; \to \mbox{ quarks (leptons) } + \:S \;.
\label{74}
\end{equation}
Such processes would appear if the weak decay Lagrangian is modified from,
say the conventional four-fermion form to the ``four-fermion + spurion''
interaction,
\begin{equation}
{\cal L}_{\rm weak}(x) \; \to \; S(x)\,{\cal L}_{\rm weak}(x)\;.
\label{75}
\end{equation}
For simplicity, it is convenient to assume, as in Eq.\,(\ref{72}) that
the spurion field does not carry spacelike momentum.

The amplitude ${\cal A}(\omega)$ has the usual analytic properties, and the
discontinuity across the physical cut at which the point $\omega=0$ is
located, describes the total decay width we are interested in. The OPE
for the inclusive widths relies on the fact that the short-distance
expansion of ${\cal A}(\omega)$ runs in $1/(\omega-E_r)$ and can be
applied near the physical point $\omega=0$ exactly as in $\rm e^+e^-$
annihilation near a positive value of $s\gg \Lam^2$. ($E_r$ denotes
energy release.) To the same extent,
in principle, a certain smearing can be required if the hadronic
probabilities still exhibit the resonance structure.

Thus, there is no theoretical peculiarity in the asymptotic applications
of the OPE for nonleptonic widths. It does not create a conceptual
difference to perform a short-distance expansion of a single quark
Green function (semileptonic widths or deep inelastic scattering), the
product of two Green functions ($\rm e^+e^-$ annihilation) or the
product of three quark Green functions (the nonleptonic widths).

Alternatively, smearing in $\omega$ can be phrased as smearing over
the interval of $m_Q$. Indeed, in the heavy quark limit the amplitudes
depend on just the combination $m_b-\omega$,
\begin{equation}
{\cal A}(\omega, m_Q) \;\simeq {\cal A}(0, m_Q-\omega)
\label{77}
\end{equation}
(there are power corrections to this relation associated with
explicit mass effects in the initial state). Therefore, in practical
terms one can phrase the smearing as an averaging over the interval of
the heavy quark mass, which may look more transparent.

After this general digression, we now return to specifically WA in
two-dimensional QCD. It is commonly accepted that, for the two-point
current correlators, both at $N_c \to \infty$ or finite $N_c$, the
properly averaged absorptive hadronic parts asymptotically coincide with
the leading OPE expression given by the free quark diagram. As was
mentioned above, for massless quarks this property holds {\em
identically} for vector and axial currents. For the scalar current the
asymptotic
correspondence in the 't Hooft model has been illustrated already in
Ref.\,\cite{callan} (for a recent discussion and earlier references, see
Ref.\,\cite{zhang}). For the WA width, however, we need  the
$m_{u,d}$-suppressed effects.  The OPE in $D=2$ yields at
$m_u\!=\!m_d\!=\!m$ (for arbitrary $N_c$)
$$
\Pi(q^2)=
N_c \! \left[
\frac{1}{q^2} \!+ \!\frac{2m^2}{q^4\sqrt{1\!-\!\frac{4m^2}{q^2}}}
\ln{\frac{\sqrt{1\!-\!\frac{4m^2}{q^2}}\!+\!1}
{\sqrt{1\!-\!\frac{4m^2}{q^2}}\!-\!1}}
\right]  -
\frac{2\pi \langle 0|{m_u\bar u u \!+ \!m_d\bar d d }|0\rangle}{q^4} +
{\cal O}\!\left(\frac{m^2 \ln{q^2}}{q^6} \right) \! ,
$$
\begin{equation}
\rho(q^2)\;=\;
N_c\,\frac{2m^2}{q^4} \frac{1}{\sqrt{1-\frac{4m^2}{q^2}}}\,+\,
2\pi \langle 0|m_u\bar u u + m_d\bar d d |0\rangle \, \delta^\prime(q^2)
\;+\;
{\cal O}\left(\frac{m^2}{q^6} \right)\;,
\label{79}
\end{equation}
where the first term in both equations is just the free quark loop.
There is little reason to doubt the OPE
for the subleading terms either. Nevertheless, it is instructive to give
here the direct derivation of the next-to-leading term $\sim
(m_u^2+m_d^2)/q^4 $ in $\rho(q^2)$ directly from the 't Hooft equation.

We follow here the approach of Ref.\,\cite{D2} based on sum rules. In
the context of the Euclidean polarization operator similar considerations
ascend to the earliest papers on the model,  Refs.\,\cite{callan,einhorn}.
To simplify the expressions, we will suppress the explicit powers of
$N_c$ which enter in a trivial way, and usually will
also omit the mass scale factor $\beta$, assuming that all
energies are measured in
units of $\beta$. Then Eqs.\,(\ref{34}),\,(\ref{35}) take the form
\begin{equation}
\Pi(q^2)\;=\;
\sum_n \,\frac{c_n^2}{q^2-M_n^2}\;, \qquad
\rho(q^2)\;=\; \sum_n \,c_n^2\,\delta\left(q^2-M_n^2\right)
\label{80}
\end{equation}
with
\begin{equation}
c_n\;=\; \int_0^1 \,{\rm d}x \:\varphi_n(x)\;.
\label{81}
\end{equation}
The completeness of eigenstates yields
\begin{equation}
\sum_n \, c_n^2\;=\;
\sum_n \,\int_0^1 \,{\rm d}x\, {\rm d}y \:
\varphi_n(x) \varphi_n(y) \;=\; 1\;.
\label{82}
\end{equation}
On the other hand, integrating the 't Hooft equation from $0$ to $1$ we
get
\begin{equation}
c_n\;=\; \int_0^1\, {\rm d}x\;\varphi_n(x) \;=\; \frac{1}{M_n^2}\,
\int_0^1\, {\rm d}x\;\left(\frac{m_d^2}{x}+ \frac{m_u^2}{1-x}\right)\,
\varphi_n(x)
\;.
\label{83}
\end{equation}
Therefore, we get the second sum rule
\begin{equation}
\sum_n \, M_n^2 c_n^2\;=\;
\sum_n \,
\int_0^1 {\rm d}x\;\varphi_n(x) \int_0^1 {\rm d}y \left(\frac{m_d^2}{y}+
\frac{m_u^2}{1\!-\!y}\right)\,
\varphi_n(y)\,=\,
\int_0^1 {\rm d}x \left(\frac{m_d^2}{x}+ \frac{m_u^2}{1\!-\!x}\right)\,.
\label{84}
\end{equation}
The integral logarithmically diverges at $x\to 0$ and $x\to 1$, which
corresponds to the behavior
\begin{equation}
\Pi(q^2)\; \stackrel{q^2 \to \infty}{\simeq} \;
\frac{1}{q^2}-\frac{m_u^2+m_d^2}{q^4}\ln{\frac{q^2}{m^2}} +
{\cal O}\left(\frac{1}{q^4}\right)
\label{85}
\end{equation}
given by the free quark diagram, Eq.\,(\ref{79}). The divergence of the sum in
Eq.\,(\ref{84}) is associated with the high excitations $n$. Therefore,
quantifying the divergence allows one to determine the asymptotic
behavior of $c_n^2$.

To render the sum in Eq.\,(\ref{84}) finite we must introduce an
ultraviolet regularization. For the logarithmic divergence the exact way
is not essential -- one is to add a hard cutoff factor
$\vartheta(\Lambda^2 \!-\! M_n^2)$. For analytic computations the Borel-type
regularization by the factor ${\rm e}^{-M_n^2/\Lambda^2}$ is usually
convenient.

For the regularized sums (we mark them with the superscript $\Lambda$)
the completeness condition is modified,
\begin{equation}
{\sum_n}^\Lambda \,\varphi_n(x) \varphi_n(y) \;=\; G(x,y; \Lambda)\;,
\label{86}
\end{equation}
and the Green function $G$ becomes a ``finite-width" $\delta$-like
distribution with the width
\begin{equation}
\Delta \;\sim\; \frac{1}{\Lambda^2} \;.
\label{87}
\end{equation}
This regularizes the sum in Eq.\,(\ref{84}):
$$
\sum_{M_n^2<\Lambda^2} \,M_n^2 c_n^2\;=\;\int_0^{\Lambda^2}\,
q^2 {\rm d}q^2\: \rho(q^2)\;=\;
\int_0^1\, {\rm d}x\,{\rm d}y
\;\left(\frac{m_d^2}{x}+ \frac{m_u^2}{1-x}\right)\,G(x,y; \Lambda)\;=
$$
\begin{equation}
=\; \left(m_u^2+m_d^2\right)\left(\ln{\Lambda^2} \:+\:
\mbox{const}\right)\;.
\label{90}
\end{equation}
One has, for instance, for the sum over an interval of highly excited
states
\begin{equation}
\int_{\Lambda_1^2}^{\Lambda_2^2}\,
q^2 {\rm d}q^2\: \rho(q^2)\;=\;
\sum_{\Lambda_1^2<M_n^2<\Lambda_2^2} \,M_n^2 c_n^2
\;=\;
\left(m_u^2+m_d^2\right)\ln{\frac{\Lambda_2^2}{\Lambda_1^2}} \:+\:
{\cal O}\left(\frac{1}{\Lambda^2}\right)
\;.
\label{91}
\end{equation}
The sum rule (\ref{90}) proves that the asymptotics of the smeared ${\rm
Im}\,\Pi(q^2)$ coincides with the free quark loop result through terms
$m^2/q^4$. It is easy to see that the nontrivial corrections in the OPE
also emerge only with higher-order terms in $1/q^2$. Note that
Eqs.\,(\ref{90}-\ref{91}) hold both for light ($m\ll \beta$) and heavy
($m\gg \beta$) quarks. However, for the asymptotics to start, the
condition $\Lambda \gg m_{u,d}$ must be observed.

Since $J_\mu^5=\epsilon_{\mu\nu}J_\nu$ and using equation of motion
$\partial_\mu J_\mu^5=(m_u\!+\!m_d) \bar{u} i\gamma_5 d$, by the same token
we showed the leading-order duality between the hadronic saturation and
the partonic expression for the absorptive part of the pseudoscalar
current. A direct derivation in the same approach is described in
Appendix\ 1.

As expected, for large $M_n$ one finds the residues $c_n \sim m_{u,d}$.
Let us
note that for light quarks $c_n$ are only linear in $m_q$: since for
light quarks $\varphi_n(x) \sim x^{m_u}$ at $x \to 0$ (and likewise at
$x \to 1$), the end points of integration in Eq.\,(\ref{83}) bring in the
$1/m_q$ enhancement.

Combining the sum rule Eq.\,(\ref{90}) with the asymptotics of the
't~Hooft eigenvalues
$$
M_n^2\;\simeq\; \beta^2 \pi^2 \,n\;,
$$
we obtain
\begin{equation}
c_n^2\;\simeq\; \pi^2 \, \beta^2 \frac{m_u^2+m_d^2}{M_n^4}\;\simeq\;
\frac{m_u^2+m_d^2}{\pi^2\,\beta^2\, n^2}
\;.
\label{94}
\end{equation}
Again, these asymptotics are valid if ``averaged'' over an interval of
$n$.

It must be noted that the explicit constant in Eq.\,(\ref{87}) is not
important. A more detailed derivation of the large-$\Lambda$ asymptotics
uses the semiclassical expansion of the 't~Hooft wavefunctions. We show
in Appendix~1 that the domain of integration where
$x<\frac{1}{\Lambda^2}$ or $y<\frac{1}{\Lambda^2}$ yields only a finite
contribution to the integral in Eq.\,(\ref{90}) (and likewise in the
vicinity of $x=1$ or $y=1$). At the same time, in the domain
$x,y\,\gg\,\frac{1}{\Lambda^2}$ the approximation
$G(x,y;\Lambda)=\delta(x-y)$ is applicable.

With the relation for the WA width Eq.\,(\ref{56}), the comparison of
Eq.\,(\ref{91}) and the OPE asymptotics Eq.\,(\ref{79}) demonstrates
that the smeared width in the 't~Hooft model coincides with the
OPE width at least through terms $m_{u,d}^2/m_Q^2$.

\section{Pauli Interference}

In this section we address the effect of interference in the weak decay
width of the heavy mesons. As explained in the Introduction, it has an
independent interest. Similar to the partonic free-quark decay width, PI
is a `subleading' $1/N_c$ effect, with $\Gamma^{\rm PI} \sim N_c$ rather
than $N_c^2$. Therefore, the expressions for the amplitudes are not as
trivial as for WA. Nevertheless, it is not difficult to demonstrate
that, again, the quark-based OPE predictions coincide with the actual
hadronic widths.

To incorporate PI we must have the flavor of the antiquark produced in
the decay of virtual $W$ coinciding with the flavor of the spectator;
we shall call it $u$. Moreover, the weak decay Lagrangian must
contain two different color structures to have PI at the same order in
$N_c$ as the free partonic width. So, we adopt, for simplicity,
\begin{equation}
{\cal L}_{\rm weak} \;=\;-\frac{G}{\sqrt{2}}\,
\left(\, a_1(\bar c \gamma_\mu b)\, (\bar{d}\gamma^\mu u )\; + \;
a_2(\bar d \gamma_\mu b)\,(\bar{c}\gamma^\mu u ) \, \right)
\; + \; {\rm H.c.}\;,
\label{95}
\end{equation}
where, again for notational transparency, we identified $Q$ with $b$ and
called $q$ by $c$.

In this case the decay width has three terms,
\begin{equation}
\Gamma_{H_Q}\;=\;\frac{G^2}{2} N_c\,
\left(\, a_1^2 \Gamma_1 +a_2^2 \Gamma_2 + 2a_1 a_2 \Gamma_{12}\right)\;,
\label{96}
\end{equation}
where $\Gamma_1$ and $\Gamma_2$ are ${\cal O}(N_c^0)$. Clearly,
$\Gamma_1(m_b, m_c,m_u,m_d)=\Gamma_2(m_b,m_d,m_u,m_c)$ holds.

The asymptotics of the non-interference width $\Gamma_{1}$ ($\Gamma_{2}$)
for the
't~Hooft model was calculated in Ref.\,\cite{D2} and shown to be given
by the OPE one. Now we address the analogous question for $\Gamma_{12}$.

The leading (in $m_Q$) contribution to the decay width described by the
free parton diagram in Fig.\,1a suggests that $\Gamma_{12} \sim 1/N_c$.
For example, for usual $V$--$A$ interaction in $D=4$ one would have
\begin{equation}
\Gamma_{12}^{\rm parton}\;=\;\frac{1}{N_c}\,\Gamma_{1}^{\rm parton}
\;=\;\frac{1}{N_c}\,\Gamma_{2}^{\rm parton}
\label{97}
\end{equation}
(the explicit factor depends on the Lorentz structure of
${\cal L}_{\rm weak}$). Such $N_c$-subleading effects are rather
complicated. This suppression, however, is not always present
\cite{ruckl}. As
discussed earlier, invoking the spectator quark through the
spectator-dependent effects like WA or PI can bring in an $N_c$-enhancement
by effectively eliminating the generic $1/N_c$ suppression of the free
quark width. As a result, at the price of a power suppression in $m_Q$
one can have the $N_c$-unsuppressed manifestation of the interference of
the two color amplitudes in ${\cal L}_{\rm weak}$,
\begin{equation}
\Gamma_{12}^{\rm (PI)}\;\sim \; {\cal O}\left(N_c^0\right)\;.
\label{98}
\end{equation}
Thus, on the one hand, studying PI allows one to address the
interference of the color amplitudes in a straightforward way relying on
the $1/N_c$ expansion. On the other hand, considering the term $\sim a_1
a_2$ in the decay width in the limit $N_c\to \infty$ automatically
singles out the power-suppressed effect of PI. This goes in contrast
with the usual situation where isolating PI formally requires subtracting the
decay width of the similar heavy flavor hadron with the spectator(s)
having the same mass but with the flavor which is sterile in weak
interactions.

The simple quark diagram describing PI is shown in Fig.\,2. To
leading order it generates the operator
$$
\hat{\Gamma}^{\rm PI}\;= \;
- 2a_1 a_2
\frac{G^2}{2K}\left\{
\left(1-\frac{m_c^2+m_d^2}{m_Q^2}\right)
(\bar b \gamma_\mu \gamma_5 u)\,(\bar u \gamma_\mu \gamma_5 b)
\;-
\right.
$$
\begin{equation}
\left.
-\;
\frac{2m_c m_d}{m_Q^2}\,
\left[(\bar b u)\,(\bar u b)\,+\,
(\bar b i\gamma_5 u)\,(\bar u i\gamma_5 b) \right]
\right\}
\;,
\label{99}
\end{equation}
with $2 m_Q K$ having the meaning of the quark spacelike momentum in the
final state:
$$
K \;= \;\left[\left(1-\frac{(m_c+m_d)^2}{m_Q^2}\right)\,
\left(1-\frac{(m_c-m_d)^2}{m_Q^2}\right)\right]^{1/2}\;.
$$
It is worth noting that this contribution is not chirally suppressed.
Therefore, it is meaningful and convenient to consider it in the
limit $m_c=m_d=0$.

For $B^-$ mesons having $\bar{u}$ spectator, the operators in
Eq.\,(\ref{99}) have the $N_c$-favorable color structure and, therefore,
their expectation values are given by vacuum factorization:
\bea
\label{100}
\frac{1}{2M_B} \langle B^-|
(\bar b \gamma_\mu \gamma_5 u)\,(\bar u \gamma_\mu \gamma_5 b)
|B^- \rangle
\; &=& \;  \frac{1}{2}\, f_B^2 M_B\\
\nonumber
\frac{1}{2M_B} \langle B^-|(\bar b u)\,
(\bar u b)\,+\,(\bar b i\gamma_5 u)\,(\bar u i\gamma_5 b) |B^- \rangle
\; &=& \; \frac{f_B^2 M_B^3}{2(m_b+m_u)^2}
\;.
\eea
In particular, at $m_c=m_d=0$ one gets
\begin{equation}
\Gamma^{\rm PI}\;= \; -2 a_1 a_2\,  \frac{G^2}{4}\; f_B^2 M_B\;.
\label{101}
\end{equation}
We note that $\Gamma^{\rm PI}$ asymptotically approaches a constant when
$m_Q \to \infty$.

It is interesting that there are no $1/m_Q$ corrections (at small
$m_{c,d}$) to the above result.
This is a peculiarity of two dimensions where the absorptive part of the
(di)quark loop in  Fig.\,2 scales as the momentum to the zeroth power
and, thus, does not depend on
whether one uses $p_b$ or $P_B$ as the momentum flowing into it. The
corrections to the Wilson coefficient as well as other higher-order operators
can induce only terms suppressed by at least two powers of inverse mass.

Let us now consider the decays in terms of hadrons. In the absence of
WA, the leading-$N_c$ final states are pairs of mesons.
The partial decay width $B\to D^0_k \pi^-_n$
takes the general form
\begin{equation}
\Gamma_{kn}\;= \; \frac{G^2}{8 M_B^2 |\vec{p}\,|}\:
\left[\,
a_1^2|{\cal A}_k {\cal B}_n|^2\, +\, a_2^2|{\cal A}_n {\cal B}_k|^2 \,+ \,
2 a_1 a_2 \,{\rm Re}\, {\cal A}_k {\cal A}_n^* {\cal B}_k^* {\cal B}_n \,
\right]\;,
\label{102}
\end{equation}
where ${\cal A}$ and ${\cal B}$
schematically denote the ``multiperipheral'' $B\to k$
transition amplitudes and the ``pointlike'' meson creation amplitudes,
respectively:
\begin{equation}
{\cal A}_k \;\sim \; \langle k|J_\mu |B \rangle\;, \qquad \;
{\cal B}_n \;\sim \; \langle n|J_\mu |0 \rangle\;.
\label{103}
\end{equation}
We denote by $\vec{p}$ the rest-frame momentum of the final state
mesons. The PI term is then given by the sum
\begin{equation}
\Gamma^{\rm PI}\;= \; 2 a_1 a_2\,\frac{G^2}{8 M_B^2}\:
\sum_{k,n}\,\frac{1}{|\vec{p}\,|}
\,{\cal A}_k {\cal A}_n^* {\cal B}_k^* {\cal B}_n
\;.
\label{104}
\end{equation}

\thispagestyle{plain}
\begin{figure}[hhh]
 \begin{center}
 \mbox{\epsfig{file=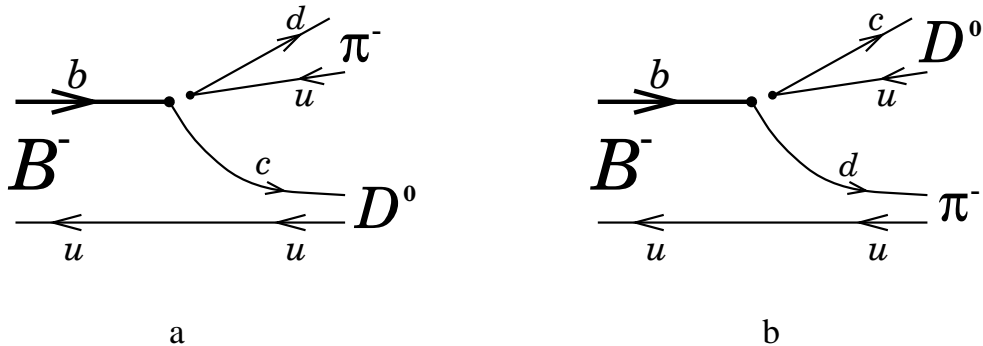,width=13cm}}
 \end{center}
 \caption{
Large-$N_c$ decay amplitudes induced by charge-current (a) and
neutral-current (b) terms in the weak decay Lagrangian.
}
\end{figure}

Both $\Gamma_1$
and $\Gamma_2$ are saturated by the final states of the type
$D^0_k \pi^-_n$ with various excitation indices $k$ and $n$. However,
the production mechanism differs: while the ``charge-current''
interaction $\sim a_1$ produces $\pi^-_n$ by the weak current
``pointlike'' and $D^0_k$ in a ``multiperipheral'' way (see Fig.\,4a),
the situation reverses for the ``neutral-current'' amplitudes
proportional to $a_2$, Fig.\,4b. These two sources of the final state
mesons have distinct features for heavy enough $Q$: the multiperipherally
produced mesons have the mass squared distributed in the interval from
$0$ to $\sim \Lam m_Q$. The bulk of the point-like produced mesons have
the mass squared \footnote{for the vector-like
current;  it would be evenly spread from about $m^2$ to $m_Q^2$ if the
weak vertex were scalar.} $\lsim m_q \Lam$ or $m_q^2$. Ref.\,\cite{D2}
demonstrated these OPE-suggested facts
explicitly in the 't~Hooft model.

As a result, interference becomes possible only at a small,
$\sim 1/m_Q$ slice of the principal decay channels. This qualitatively
explains
the $1/m_Q$ power suppression of PI which is automatic in the OPE.

We will now demonstrate the quantitative matching between the OPE-based
calculation and the hadronic saturation of the interference width. To
make the proof most transparent, we start with the simplest possible
case when all final state quarks $u,d,c$ are massless. While not
affecting the OPE analysis, this limit significantly simplifies the
expressions for the individual hadronic amplitudes, as explained in
Ref.\,\cite{D2}. In the case at hand, for example, only $n=0$ survives
for the decay amplitude $\sim a_1$ (Fig.\,4a) and $k=0$ for the
amplitude $\sim a_2$ (Fig.\,4b). The interference then resides in the
single final state
containing the lowest lying massless $D^0$ and
$\pi^-$.
Moreover, the corresponding transition amplitudes ${\cal A}$
between two mesons take
particularly simple form at $q^2=0$ in terms of their 't~Hooft
wavefunctions \cite{D2}:
\begin{equation}
q_\mu\: \frac{1}{2M_B} \langle k|\epsilon_{\mu\nu} J_\nu |B \rangle \;=\;
-q_z\, \int_0^1 \, {\rm d}x \: \varphi_k(x) \varphi_B(x)
\;,
\label{105}
\end{equation}
where we have recalled that
${\cal B}_0\sim i\epsilon_{\mu\nu} f_0 P_\nu^{(0)}$, and
therefore considered only
the relevant light-cone component of the amplitude. Then
we have
\begin{equation}
\Gamma^{\rm PI}\;= \; -2 a_1 a_2\,\frac{G^2 M_B N_c}{4\pi}\:
\left|\int_0^1 \, {\rm d}x \: \varphi_B(x)\right|^2\;= \;-
2 a_1 a_2\,\frac{G^2}{4} \, f_B^2 M_B
\label{106}
\end{equation}
(we have used the fact that $\varphi_0(x)=1$,
$f_0=\sqrt{N_c/\pi}$ for massless quarks).
The minus sign emerges since the direction of the vector playing the
role of $\vec{q}$ is
opposite for the two interfering amplitudes.

Thus, the OPE asymptotics Eq.\,(\ref{101}) is exactly reproduced.
Apparently, there is no violation of local duality {\it at all} for PI
in the case $m_u=m_d=m_c=0$! This is not surprising -- in this limit the
only threshold in $\Gamma^{\rm PI}$ occurs at zero mass, and the OPE
series can have the same convergent properties in Minkowskian as in
Euclidean space.

With $m_{u,d,c}\ne 0$ the interference effects are saturated
by several final state pairs of mesons, even if the masses are
small compared to $m_Q$. It is still not difficult, though, to check
that the leading OPE term  Eq.\,(\ref{101}) is reproduced. We keep in
mind that at nonzero masses the width exhibits the threshold
singularities due to the singular two-body phase space $1/|\vec{p}\,|$ in
$D=2$. Since it is integrable, the threshold spikes do not affect the
width smeared over the interval of mass $\Delta m_Q \sim 1/m_Q$.

The idea of the proof is suggested by the detailed kinematic duality
between the partonic and hadronic probabilities. The bulk of
pointlike-produced mesons have masses squared $M_n^2$ not exceeding
$\beta^2$ or
$m_q^2$, while for multiperipherally-created mesons $k$ this scale is
$\sim \beta m_Q$ or $m_{\rm sp} m_Q$. More precisely \cite{D2},
for the decay rates $\sim a_1^2$ (Fig.\,4a)
\bea
\frac{1}{\Gamma_{\rm tot}}\:\sum_{k} \,\sum_{M_n>{\rm const}\cdot m_Q}
\Gamma_{kn} \;& \propto \; & \frac{m_{u,d}^2}{m_Q^2}\:,\\
\frac{1}{\Gamma_{\rm tot}}\:\sum_{M_n\ll m_Q} \,
\sum_{M_k>{\rm const}\cdot m_Q}
\Gamma_{kn} \;&\propto \; &\frac{1}{m_Q^5}
\;.
\label{107}
\eea
Then, calculating the width we can expand around the free quark
kinematics $M_n=M_k=0$. In particular, we set
\begin{equation}
\frac{1}{|\vec{p}\,|}\; =\;\frac{2}{M_B} \left(
1+\frac{M_k^2+M_n^2}{M_B^2}\,+\,...
\right)\;.
\label{108}
\end{equation}
Additionally, we can expand the transition formfactors in amplitudes
${\cal A}_{k}$ in $q^2$:
\begin{equation}
{\cal A}_k(M_n^2)\;\simeq\; {\cal A}_k(0) \:+\:\frac{M_n^2}{m_Q^2}\:
m_Q^2 \!\left.\frac{{\rm d}{\cal A}_k}{{\rm d}q^2}\right|_{q^2=0}\;,
\label{109}
\end{equation}
and likewise for ${\cal A}_n(M_k^2)$. In factoring out $1/m_Q^2$ in the slope
of the amplitude we accounted for the fact that it scales as $1/m_Q^2$
in this kinematics. Indeed, the $t$-channel resonances have masses
exceeding $m_Q$, and
the kinematics (the fractions of the
light-cone momenta entering computation of the transition
amplitudes, see the next section and Appendix\,\,2) likewise
depend on $q^2$ only as $q^2/m_Q^2$.

To obtain $\Gamma^{\rm PI}$ with an accuracy $1/m_Q^2$ of the free quark
width, we actually expand the particular two-body decay amplitude only
in $q^2$, that is, do not neglect $M_k^2$-dependence for the amplitude
$\sim a_1$ or $M_n^2$-dependence for the amplitude which is proportional
to $a_2$. The expressions for the decay amplitudes at $q^2=0$ are very
simple \cite{D2}:
$$
{\cal M}_{kn}\;=\; G \sqrt{\frac{N_c}{2\pi}} \,\left(
a_1 {\cal M}_{kn}^{(1)} \,+\,a_2 {\cal M}_{kn}^{(2)}
\right)\;,
$$
\bea
\nonumber
&{\cal M}_{kn}^{(1)} \;=\; &(M_B^2-M_k^2)\:
\int_0^1 \, {\rm d}x \: \varphi_B(x) \varphi_k(x)\cdot
\int_0^1 \, {\rm d}y \: \varphi_n(y)\\
&{\cal M}_{kn}^{(2)} \;= -&(M_B^2-M_n^2)\:
\int_0^1 \, {\rm d}x \: \varphi_B(x) \varphi_n(x)\cdot
\int_0^1 \, {\rm d}y \: \varphi_k(y)\;,
\label{110}
\eea
$$
2|\vec{p}\,|M_B \;\simeq\; M_B^2-M_k^2-M_n^2\;.
$$
Then we have
$$
\sum_{k,n}  \!\Gamma^{\rm PI}_{kn} \! \simeq \!
- \!2a_1 \!a_2 \!
\frac{G^2\! M_B\! N_c}{4\pi}
\sum_{k,n} \! \int_0^1 \! {\rm d}x \, \varphi_{B}(x) \varphi_k(x)
\int_0^1 \!{\rm d}y \, \varphi_B(y) \varphi_n(y)
\int_0^1  \! {\rm d}z \, \varphi_n(z)
\int_0^1  \! {\rm d}t \,\varphi_k(t)
$$
\begin{equation}
=\; -2a_1 a_2\, \frac{G^2 M_B N_c}{4\pi M_B^2}
\left| \int_0^1 \, {\rm d}x \: \varphi_B(x) \right|^2
\;,
\label{111}
\end{equation}
or
\begin{equation}
\Gamma^{\rm PI}\;= \; -2 a_1 a_2\, \frac{G^2}{4}\; f_B^2
\,M_B\;.
\label{112}
\end{equation}
We extended summation over $k$ and $l$ in Eq.\,(\ref{111}) to include all
states, since the contribution of additional, kinematically forbidden meson
pairs is suppressed by high powers of $1/m_Q$.

This expression is valid up to the relative $1/m_Q^2$ corrections.
Indeed, the leftover effect of the slope of the transition formfactor
$F_{Bk}$ is
quadratic in $m_q/m_Q$. For example, using representation
Eq.\,(\ref{83}) we obtain a sum rule which allows one to cast it in the
form
$$
\sum_{n}\, \delta\, {\cal M}_{kn}^{(1)} \cdot {\cal M}_{kn}^{(2)}
\;=\;
\sum_{n}\,
\frac{\partial}{\partial q^2}F_{Bk} \cdot M_n^2\,c_n \cdot
\int_0^1 \, {\rm d}y \: \varphi_B(y) \varphi_n(y)\cdot
c_k \;=
$$
\begin{equation}
=\;
c_k\:\frac{\partial}{\partial q^2} F_{Bk} \cdot
\int_0^1 \, {\rm d}y \: \varphi_B(y) \left(
\frac{m_d^2}{y}+\frac{m_u^2}{1-y}
\right)
\label{113}
\end{equation}
(and likewise for $\delta\, {\cal M}_{kn}^{(2)}$). The convergence of
the integral over $y$ shows that this effect is saturated at small $n$
and is of order
$\frac{m_q^2}{m_Q^2}$ or $\frac{m_q\beta}{m_Q^2}$, whichever is larger.

A more accurate consideration reveals that the two amplitudes in
Eqs.\,(\ref{110}) have the factors $(\!-1\!)^n$ and $(\!-1\!)^k$,
respectively, and their product, additionally, the factor
$(\!-1\!)^{({\rm P_{H_Q}} +n+k)}$. (The latter is related to the opposite
direction of ``$\vec q\,$'' in the two amplitudes
and is readily understood since this is a
parity-conserving decay $H_Q\to k+n$ with the meson parities
${\rm P_{H_Q}}$,
$(\!-1\!)^n$ and $(\!-1\!)^k$.) Therefore, the sign of $\Gamma^{\rm PI}$
is given by the parity of $H_Q$, which is manifest for the OPE result in
$D\!=\!2$, cf.\ Eq.\,(\ref{100}).

Thus, we see that $\Gamma^{\rm PI}$ agrees with the expression given by
the free quark loop just to the accuracy suggested by the OPE.

It is not difficult to estimate the effects of violation of local
duality in PI related to the thresholds, for small but nonvanishing $m_q$.
Since the two-body phase space is singular, different ways to
gauge its strength will yield different power of its asymptotic
suppression.
Full information is just given by the nature of the threshold
singularity, the scaling of the corresponding residues and the
asymptotic distance between the principal thresholds.
This would show the contribution to PI of any new decay channel, close to
the mass where it opens, where the corresponding width is not literally
given by the OPE.

It turns out that the magnitude of local duality violation in PI
essentially depends on the relation between the final state masses. The
strongest effect comes from the kinematics where one of the mesons belongs
to low excitations while another has the large mass close to $m_b$.

The case when $m_u=m_d=0$ (but $m_c > 0$) is somewhat special. Here one
of the interfering decay amplitudes vanishes at the thresholds, and
$\Gamma^{\rm PI}$ simply experiences a finite jump:
\begin{equation}
|\delta \Gamma^{\rm PI}_{k0}|
\;\simeq\;
\frac{G^2}{2}\, 2a_1a_2 \cdot {\rm const}\, f_B \,m_c
\frac{\beta^{9/2}}{M_{H_Q}^{9/2}}\: \vartheta (M_{H_Q}\!-\!M_k)
\;,
\label{115}
\end{equation}
$$
M_{k+1}-M_k\simeq \frac{\pi^2\beta^2}{2M_{H_Q}}\;.
$$
Here we used the semiclassical calculations of the transitions to highly
excites states given in Sect.\,3, Eq.\,(\ref{94}) and in
Ref.\,\cite{D2}, Eq.\,(79). The latter estimate for the $H_Q\to D^{k}$
transition amplitude is valid up to a factor of order one; accepting it
at face value would yield $\sqrt{3\pi}$ for the constant in
Eq.\,(\ref{115}).

The referred asymptotics determined the absolute magnitude of the decay
amplitudes relevant for usual decay probabilities, but not their sign
which plays a role in interference which can be both constructive and
destructive. A more careful analysis suggests that the relative sign of
the two amplitudes alternates for successive thresholds. Therefore, at
$m_u\simeq m_d\ll \beta$ and $m_c\ll m_Q$ we have the following ansatz:
\begin{equation}
\delta \Gamma^{\rm PI}_{\rm osc}
\;\simeq\;
\frac{G^2}{2}\, 2a_1a_2 \cdot {\rm const}\, \sqrt{3\pi} \,f_B \,m_c
\frac{\beta^{9/2}}{M_{H_Q}^{9/2}}\: \sum_k (\!-\!1)^k \,
\vartheta \left(M_{H_Q}\!-\!\pi\beta\sqrt{k}\right)
\;.
\label{117}
\end{equation}
The amplitude of oscillations in PI scales down at least as $1/m_Q^5$.
At large $m_Q$ the threshold widths are much smaller than even the
individual principal widths saturating $\Gamma^{\rm PI}$ at
$m_{c,u,d}\ne 0$.

When $m_{u,d}$ are nonzero, the picture changes essentially in two
respects. First, neither decay amplitude vanishes at the threshold,
since the two-momentum of the lighter meson does not vanish: $q_0
\simeq M_1$ instead of $q_0= \mvec{q} \simeq M_{H_Q}\!-\! M_{\rm thr}$
if
$q^2=0$. Second, the phase space factor $1/\mvec{p}$ becomes now
$[2M_1(M_{H_Q}\!-\!M_{\rm thr})]^{-1/2}$ vs.\ $1/(M_{H_Q}\!-\!M_{\rm
thr})$ for $M_1=0$. ($M_1$ is the mass of the lighter meson and its
momentum is called $q$ here. We assume
that $M_1$ is much larger than the resonance spacing $\sim
\beta^2/m_Q$.) Otherwise, the scaling of the transition amplitudes
remains the same. Therefore, in this case we have
\begin{equation}
\delta \Gamma^{\rm PI}_{\rm osc}
\;\propto \;
\frac{G^2}{2}\, 2a_1a_2\:
m_c m_\pi^{1/2}\,
\frac{\beta^{5}}{M_{H_Q}^{5}}\: \sum_k (\!-\!1)^k \,
\frac{\vartheta \left(M_{H_Q}\!-\!M_{\rm thr}^{(k)}\right)}
{\sqrt{M_{H_Q}\!-\!M_{\rm thr}^{(k)}}}
\;,
\label{118}
\end{equation}
$$
M_{\rm thr}^{(k)}\;\simeq\; m_\pi + \pi \beta\sqrt{k} \;.
$$

Strictly speaking, at $m_{u,d} \gsim \beta$ additional light meson
states contribute, and the pattern of the threshold spikes in
$\Gamma^{\rm PI}$ becomes less even reflecting the superposition
of a number of similar structures. Additionally, at $m_u \sim m_c$ the
individual sign-alternating behavior becomes more complicated.

In principle, with all final-state masses not vanishing, there
are thresholds corresponding
to decays where both final-state mesons have masses constituting a
finite fraction of $m_Q$. The threshold amplitudes for such decays,
however, are too strongly suppressed, since both interfering amplitudes
have the chiral and the formfactor suppression:
$$
|{\cal M}_{kn}|^2_{\rm PI} \;\propto\;
\frac{m_q^2 \beta^9}{m_Q^7}\qquad \mbox { at } k,n \sim
\frac{m_Q^2}{\beta^2}\;.
$$
The phase space for such decays is also smaller since $\mvec{p} \sim
\sqrt{m_Q} (M_{H_Q}\!-\!M_{\rm thr})^{1/2}$. Additionally, these thresholds are
spaced very closely, at distances scaling as $\beta^4/m_Q^3$. Therefore,
they are subdominant for duality violation. They are related to
the subseries of
the OPE terms which appear only beyond the tree-level perturbative
computations.
\vspace*{.35cm}

To conclude this section, let us describe the physical picture which
emerges from the analysis. In particular, we can see how the
interpretation problem
mentioned in the Introduction is resolved. As expected, the explicit
analysis yielded nothing about colored diquark
correlator, directly.
Instead, we observe the duality of differently combined
quark-antiquark pairs to the hadronic states: one energetic quark ($c$ or
$d$) is to be combined with the `wee' spectator antiquark or slow
$\bar{u}$ produced in the weak vertex. It is the pair of quarks picked
up from the different final state mesons that corresponds to the large
invariant mass in the quark diagram. The completeness of the hadronic
states -- or, in other words, the duality between the parton-level and
mesonic states -- is achieved already for a single fast moving decay quark
when it picks up a slow spectator. In particular, the `hardness' of
these processes determining the applicability of the quasifree
approximation, is governed by the energy of the fast quark rather than
by the invariant mass of the pair.

There is nothing wrong with considering the colored diquark loop as
nearly free. Since the overall color is conserved in the perturbative
diagrams, in the full graph for the meson decay which would include
explicitly propagation of the spectator, there is always a color mate
for any quark in the ``partonic'' part of the diagram. Moreover,
if the leading-$N_c$ contribution is considered, such a color pairing
({\it i.e.},  which pair must be embodied into a meson) is unambiguous.

Of course, the invariant {\em mass} of a single, even fast on-shell
quark vanishes. However, this does not make the inclusive probability
for it to hadronize by picking up spectator and forming a meson, a
``soft'' quantity.
For it is not the invariant mass but
the (rest frame) momentum that determines the hardness. Indeed, the
color of the initial static heavy quark $Q$ is compensated by the slow
spectator. This initial distribution of the color field marks the rest
frame and makes the hardness parameter for the total probability to look
non-invariant if the final state is considered perturbatively as a pair of
free partons.

\section{Total (spectator-free) width through $1/m_Q^2$}

The inclusive decay widths of heavy hadrons in the 't~Hooft
model in the absence of the flavor-dependent spectator effects were
considered in detail in Ref.\,\cite{D2}. It was demonstrated
that the analytic summation of the widths for the accessible two-body
modes reproduces the $1/m_Q$ expansion of the widths in the OPE, at least
through the terms high enough in $1/m_Q$. In particular, the hadronic
width does not have any $1/m_Q$ correction which would not be present
in the OPE.

The analysis was performed for arbitrary $m_c$ and $m_{\rm sp}$, but
simplified significantly when $m_u=m_d=0$ was set (in the notations of
the present paper). Since there is little doubt that the dependence of
the hadronic
width on $m_u$ and $m_d$ is suppressed by at least two powers of $1/m_Q$,
this
simplification cannot affect the conclusion regarding possible non-OPE
$1/m_Q$ terms in the width. Nevertheless, we find it instructive to
describe
the direct computation of the terms $\sim m_{u,d}^2/m_Q^2$ in the width
based on the 't~Hooft eigenstate problem, following the approach of
Ref.\,\cite{D2} and the analysis of the previous sections.\footnote{The
consideration below was elaborated while working on paper
\cite{D2}, but was not included in the final version for the sake of
brevity.}
In particular, it illustrates that the case of nonzero masses is not any
different from $m_u\!=\!m_d=0$.

As before, we assume for simplicity that $m_u=m_d$, so that the
$\bar{u}\gamma_\mu d$ current is strictly conserved, and represent the
large-$N_c$ nonleptonic decay width as an integral of the differential
semileptonic width $\Gamma_{\rm sl}(q^2)$ over $q^2$ weighted with the
spectral density $\rho(q^2)$, Eq.\,(\ref{20}). The upper limit of
integration $q^2_{\rm max}$ comes from vanishing of $\Gamma_{\rm sl}(q^2)$
at $q^2> (M_{H_Q} \!-\! M^D_0)^2$.

$\Gamma_{\rm sl}(0)$ was analytically calculated in Ref.\,\cite{D2}:
$$
\Gamma_{\rm sl}(0)\;=\; \frac{G^2}{4\pi} \frac{m_Q^2-m_q^2}{m_Q}
\, \left[ \frac{m_Q}{M_{H_Q}}\int _0^1
\frac{{\rm d}x}{x}
\,\varphi _{H_Q}^2(x) + {\cal O}\left(\frac{1}{m_Q^5}\right) \right]\;=
$$
\begin{equation}
=\; \frac{G^2}{4\pi} \frac{m_Q^2-m_q^2}{m_Q}\, \left[
\frac{\langle H_Q|\chi _Q^{\dagger}
\frac{m_Q}{i\partial _-}\chi _Q | H_Q \rangle}{2M_{H_Q}}
+ {\cal O}\left(\frac{1}{m_Q^5}\right)
\right]\;,
\label{123}
\end{equation}
which coincides with its OPE expansion.

On the other hand, $\rho(q^2)$ does not vanish at $q^2>0$ only due to nonzero
$m_{u,d}$. Therefore, following the approach of the previous section, we expand
$\rho(q^2)$ in $m_q^2/q^2$ at large $q^2$ and, simultaneously,
$\Gamma_{\rm sl}(q^2)$ in $q^2/m_Q^2$ at $q^2=0$. To this end we
write the width as
\begin{equation}
\Gamma_{H_Q} = \int_0^{q^2_{\rm max}} \! {\rm d}q^2 \, \rho(q^2)\,
\Gamma_{\rm sl}(q^2)
 =
\Gamma_{\rm sl}(0)\,\int_0^{\infty} \! {\rm d}q^2 \, \rho(q^2) +
\int_0^{\infty} \! {\rm d}q^2 \, \rho(q^2) \left(\Gamma_{\rm sl}(q^2)
- \Gamma_{\rm sl}(0)\right) .
\label{124}
\end{equation}
Since at $q^2\sim m_Q^2$ the spectral density $\rho(q^2)$ is
explicitly proportional to $m_q^2/m_Q^2$, in the second integral we can
use for $\Gamma_{\rm sl}(q^2)- \Gamma_{\rm sl}(0)$
its leading-order approximation. With
$\int {\rm d}q^2 \: \rho(q^2)=N_c$, the first term {\em exactly} reproduces
the corresponding term in the OPE, Eq.\,(\ref{123}).

All transition formfactors are expandable in $q^2/m_Q^2$ (except,
possibly, the point $q^2\to m_Q^2$). Therefore,
the smeared width $\Gamma_{\rm
sl}(q^2)$ is likewise expandable in $q^2/m_Q^2$, and so is
$\Gamma_{H_Q}$ -- at least up to small
corrections $\sim 1/m_Q^3$ associated with the domain of $q^2$ close to
$m_Q^2$. Thus, the second term scales only as $m_{u,d}^2/m_Q^2$.

In order to calculate this term, we can use the following facts regarding the
smeared width:

$\bullet$ The exact `semileptonic' width  $\Gamma_{\rm sl}(q^2)$
coincides with the free width $\Gamma_{\rm sl}^{\rm tree}(q^2)$ to the
leading order in $1/m_Q$ when $m_Q-\sqrt{q^2} \gg \beta$.

$\bullet$ The smeared $\Gamma_{\rm sl}(q^2)$ is an analytic function of
$q^2$  and is expandable in $1/(m_Q-\sqrt{q^2})$.

$\bullet$ The smeared $\Gamma_{\rm sl}(q^2)$ does not blow up at
$q^2\to m_Q^2$.

\noindent
We shall comment on them below. Accepting these three facts for
now, and neglecting
$m_q^2$ compared to $m_Q^2$ we find
\begin{equation}
\Gamma_{\rm sl}^{\rm tree}(q^2)\;=\;\frac{G^2}{4\pi}\, m_Q\,
\vartheta(m_Q^2\!-\!q^2)\;,
\label{127}
\end{equation}
and
$$
\int_0^{\infty} {\rm d}q^2 \: \rho(q^2) \left(\Gamma_{\rm
sl}(q^2) \!- \!\Gamma_{\rm sl}(0)\right)\, =\,
\int_0^{\infty} {\rm d}q^2 \: \left(\Gamma_{\rm
sl}(q^2) \!-\! \Gamma_{\rm sl}(0)\right)\cdot \frac{1}{q^4}
\frac{{\rm d}}{{\rm d} \ln{q^2}}\,
\int_0^{q^2} \! {\rm d}t \; t \,\rho(t)
$$
\begin{equation}
=\;-\frac{G^2 m_Q N_c}{4\pi}\,
\int_{m_Q^2}^{\infty}\,\frac{{\rm d}q^2}{q^4}\,(m_u^2+m_d^2)\;=\;
-N_c \frac{m_u^2+m_d^2}{m_Q^2}\: \Gamma_{\rm sl}(0)\;.
\label{128}
\end{equation}
Here we have used the sum rules Eqs.\,(\ref{90})-(\ref{91}).

Thus, we get through order $1/m_Q^2$
\begin{equation}
\Gamma_{H_Q}\; =\;-\frac{G^2}{4\pi}\,N_c\, \frac{m_Q^2-
m_c^2-m_u^2-m_d^2}{m_Q}\,
\left(1+\frac{\beta^2}{m_Q^2}-\frac{\mu_\pi^2}{2m_Q^2}\right)
\; +\; {\cal O}\left(\frac{\beta^3\!, m_q^3}{m_Q^2}\right)\, ,
\label{130}
\end{equation}
where the kinetic expectation value
\begin{equation}
\mu_\pi^2\;=\; m_Q^2\left[
\int_0^1 \, {\rm d}x \, x^2 \varphi_{H_Q}^2(x)\,-\,
\left( \int_0^1 \, {\rm d}x \, x\, \varphi_{H_Q}^2(x) \right)^2
\right]
\label{131}
\end{equation}
represents the low-momentum part of the $1/m_Q$ expansion of the integral
in Eq.\,(\ref{123}), whereas the term $\beta^2/m_Q^2$ accounts for its
``hard'' part \cite{D2}.
The OPE result is thus reproduced explicitly.

Neglecting $m_c$ compared to $m_Q$ was essential in the above
computation. For, at
$m_c\ne 0$ the unique {\em linear} in $m_{u,d}$ effect appears which was
calculated in Ref.\,\cite{D2}, Sect.VI.A\,:
\begin{equation}
\frac{\Delta\Gamma_{H_Q}}{\Gamma_{H_Q}}\; =\;-\frac{4\pi}{N_c}\,
\langle 0|{m_u\bar u u + m_d\bar d d }|0\rangle
\frac{m_Q m_c}{(m_Q^2 -m_c^2)^2}\;.
\label{132}
\end{equation}
Of course, it has the direct OPE counterpart, see Figs.\,5. At $m_c\sim
\beta$ it scales as $\frac{\beta^2(|m_u|+|m_d|)}{m_Q^3}$.

\thispagestyle{plain}
\begin{figure}[hhh]
 \begin{center}
 \mbox{\epsfig{file=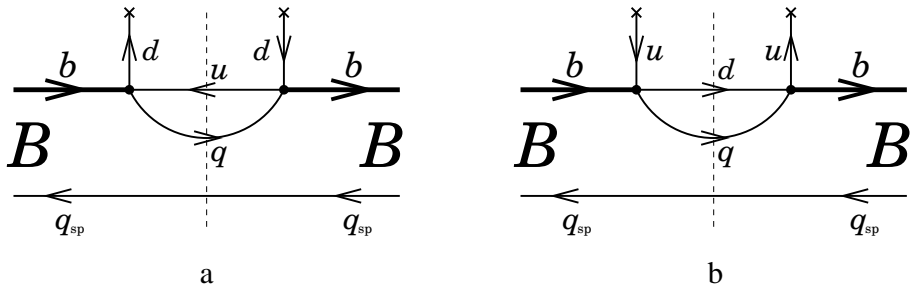,width=12cm}}
 \end{center}
 \caption{
Diagrams responsible for the linear in $m_{u,d}$ corrections to
$\Gamma_{H_Q}$.
}
\end{figure}

It is straightforward to obtain the $1/m_Q^3$ term in $\Gamma_{H_Q}$
as well. The least
trivial corrections not related to $m_u,\:m_d$ are all incorporated in
Eq.\,(\ref{123}). The only remaining part is the $1/m_Q$ term in the
explicit $\frac{m_u^2+m_d^2}{m_Q^2}$ correction.
It is associated with the domain of maximal $q^2$, {\it i.e.}
$m_Q-\sqrt{q^2}\sim \, m_c, \beta$. It has a logarithmic enhancement,
$\frac{(m_u^2+m_d^2)m_c}{m_Q^3}\ln{\frac{m_Q}{m_c}}$ coming from the
domain $m_c \ll m_Q-\sqrt{q^2} \ll m_Q$ (and, therefore, this $\log$
conforms with the one in the free quark phase space). On the other
hand, at non-equal $m_u$ and $m_d$ and light $c$ and $q_{\rm sp}$
the smeared width
contains the chiral $\log$ of the
form $\frac{(m_u-m_d)^2\beta}{m_Q^3}
\ln{\frac{\beta}{m_c+m_{\rm sp}}}$, from the domain
$\sqrt{\beta(m_c+m_{\rm sp})} < m_Q-\sqrt{q^2} < \beta$. These
contributions are given by the corresponding four-quark expectation
values
$ \langle H_Q|\bar Q \Gamma_\mu c\, \bar c \Gamma_\nu Q|H_Q\rangle$, in
agreement with the general proof of Ref.\,\cite{WA}.\footnote{At
$m_u\!=\!m_d$
these expectation values are multiplied by
$\delta_{\mu\nu} -v_\mu v_\nu$ with $v=P_{H_Q}/M_{H_Q}$, containing only
the spacelike component, and then
the contribution of the lowest pseudoscalar, ``pion'' intermediate state
comes proportional to its momentum. From the point of view of hadronic
decay modes, the pion threshold decay amplitude likewise
vanishes at $m_u\!=\!m_d$ since, for the conserved
vector current, only the current-meson
couplings $\sim \epsilon_{\mu\alpha}$ survive, and the decay amplitudes
into the mesons with the same parity as $H_Q$ vanish at the threshold.
These suppressions are eliminated when $m_u \! \ne \! m_d$, and the chiral
$\log$ appears both in the four-quark expectation value due to the pion
contribution in the timelike component, and in the smeared width due to
the pion phase space $\sim 1/(M_{H_Q}\!-\!M_{\rm thr})$.}
However, these operators being $N_c$-subleading (the expectation
values scaling as $N_c$ rather than $N_c^2$, unless WA is possible),
their values are anyway expressed in a rather ugly way in terms of
various 't~Hooft eigenfunctions both in the $s$ and $t$ channel.
Calculating this maximal-$q^2$ effect requires straightforward $1/m_Q$
expansion of the formfactors and wavefunctions, which is not too
instructive. This is briefly outlined in Appendix~2.

Now it is time to comment on the assumptions used in deriving the
correction $-\frac{m_u^2+m_d^2}{m_Q^2}$. The first one was related to the
semileptonic width at nonzero $q^2$. Of course, even the stronger
statement holds: $\Gamma_{\rm sl}(q^2)$ coincides with
$\Gamma_{\rm sl}^{\rm tree}(q^2)$ up to the terms
$\sim \beta^2/(m_Q-\sqrt{q^2})^2$. Proving this goes along the
same explicit $1/m_Q$ expansion of the wavefunctions and the formfactors
\cite{D2}. Namely, to order $1/m_Q$ one still has the transition
formfactors determined only by the overlap of the initial $\varphi_{H_Q}$
and the final $\varphi_k$ wavefunctions. The only difference is that
$q_-\ne 0$ at $q^2\ne 0$ and, as a result, the arguments of the
wavefunctions change. This purely kinematic modification accounts for
all changes to order $1/m_Q$. Some technicalities are given in
Appendix~2.

The situation changes at order $1/m_Q^2$. The peculiarity of the point
$q^2\!=\!0$ is
that the vertices do not renormalize at $q^2\!=\!0$. More precisely,
in the light-cone gauge the vertex corrections in Fig.\,6a are proportional
to $q_-$ and thus vanish, to all orders, at $q^2\!=\!0$ \cite{D2}. The only
surviving contribution is the effect of renormalization of the external
quark legs, Fig.\,6b
which is readily summed up to all orders. In the light-cone
formalism it consists in overall (IR divergent) shift in the reference
point for the light-cone energy $p_+$, and the dispersion term
$-\beta^2/(2p_-)$ formally coinciding with the replacement $m^2\, \to\,
m^2\!-\!\beta^2$ for all quark flavors.

\thispagestyle{plain}
\begin{figure}[hhh]
 \begin{center}
 \mbox{\epsfig{file=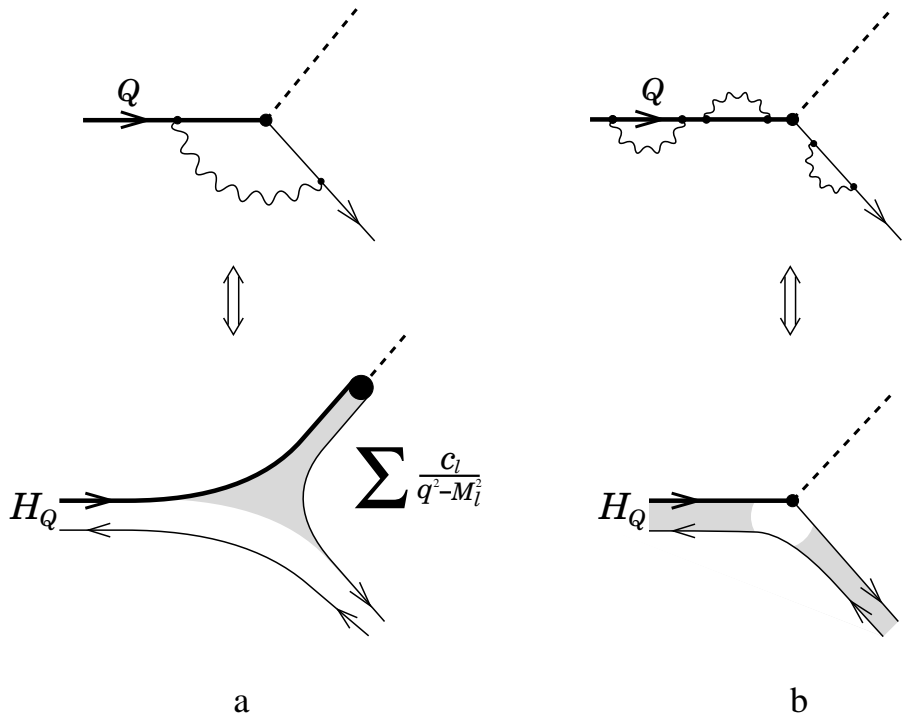,width=12cm}}
 \end{center}
 \caption{
Vertex (a) and external quark leg (b) renormalizations of weak decay
amplitudes, and their hadronic counterparts.
}
\end{figure}

This perturbative computation has an exact analogy for the actual meson
formfactors. The absence of the vertex corrections means the absence
of the $t$-channel resonance contributions at $q_-=0$. This fact is
directly seen in the explicit expressions for the formfactors in
the kinematics where
the fraction of the total light-cone momentum $\omega$ corresponding to
the momentum transfer $q$, goes to zero. All the strong interaction
effects occur in the initial and the final states and are described by
using the exact eigenfunctions instead of the plane waves.

At $q_-\ne 0$ the vertex corrections appear to order $g_s^2$, and already
the first $g_s^2$ correction in the coefficient function of the operator
$\bar{Q}Q$ becomes different from $(1-\beta^2/m_Q^2)^{-1/2}$. The vertex
corrections are dual to the $t$-channel resonance contribution in the
exact amplitude, and they also appear explicitly with the factors $q_-$
and $\beta^2$. Additionally, the original tree-level overlap is modified
both kinematically and due to the additional terms in the expansion of
the $\bar{Q}\gamma_\mu q$ current in terms of the light-cone spinors.
All this is directly observed in the explicit expressions for the meson
transition formfactors.

Regarding the two other assumptions, the fact that the transition
formfactors are regular functions of $q^2$ can be seen, of course, from
their most general analytic expressions. Another way to visualize
this is to use the approximate scaling behavior
\begin{equation}
\frac{1}{\sqrt{2m_Q}}\, \langle k|J_\mu (q) |H_Q(m_Q)\rangle \;\simeq\;
\frac{1}{\sqrt{2m_Q'}}\, \langle k|J_\mu (q') |H_Q(m_Q')\rangle
\;,
\label{134}
\end{equation}
where $q'$ is adjusted in respect to $m_Q'$ to have the same
rest-frame momentum of the final-state meson $k$:
\begin{equation}
{q'}^2\;=\; \frac{M_{H_Q}(m_Q)}{M_{H_Q}(m_Q')} q^2 \,+ \,
M_{H_Q}(m_Q')\left(M_{H_Q}(m_Q')\!-\!M_{H_Q}(m_Q)\right) \,+\,
M_k^2\left(1\!-\!\frac{M_{H_Q}(m_Q)}{M_{H_Q}(m_Q')}\right)\,.
\label{135}
\end{equation}
This freedom can be used, say, to make momentum transfer $q$
light-like,
\begin{equation}
m_Q'\;\simeq \; m_Q\;-\; \left[\frac{m_Q^2-M_k^2+q^2}{2m_Q} -
\sqrt{\left(\frac{m_Q^2- M_k^2- q^2}{2m_Q}\right)^2+
\frac{q^2 M_k^2}{m_Q^2}
} \;\right]
\label{136}
\end{equation}
and then use the exact
representation of the amplitudes {\it via} the simple wavefunctions overlaps.
This trick allows one, for example, to see the strong suppression of the
transition amplitudes to highly excited states $M_k^2 \sim m_Q^2 \gg
\beta m_Q$ at non-zero $q^2$ as well, without going into details
described in Appendix~2.
The relation Eq.\,(\ref{134}) is violated by the $\beta^{2n}$ radiative
corrections and by the subleading $1/m_Q$ terms in the expansion of
the current $J_\mu$, which are both power suppressed here.
It can be derived directly from the explicit solution of the model, see
Sects.\,2 and 4 of Appendix~2.

\section{Comments on the literature}

Two recent papers \cite{gl,gl2} claimed to have established
inapplicability of the OPE for the heavy flavor widths considering the
solvable 't~Hooft model. Ref.\,\cite{gl} addressed the conventional
spectator-independent decay channels. Numerically evaluating the
possible two-body decay rates for the values of $m_Q$ up to $m_Q=14$, it
found a small systematic excess of the decay width over the free quark
diagram which was fitted as
\begin{equation}
\frac{\Gamma_{H_Q}-\Gamma_Q}{\Gamma_Q}\;\sim \; 0.15 \frac{\beta}{m_Q}\;.
\label{140}
\end{equation}
This was regarded as the demonstration of ({\it a priori} proclaimed)
non-existence of the OPE for the nonleptonic widths. Since in the
large-$N_c$ limit the difference between the nonleptonic and
semileptonic widths disappears, this -- if true -- would mandate the
same absence of the OPE for the semileptonic widths as well, an obvious
fact ignored by the authors.

In contrast, Ref.\,\cite{D2} accomplished the analytic summation
of the large-$m_Q$ width in the 't~Hooft model, and no deviation from
the OPE was found to the high enough orders in $1/m_Q$ (the exact power
addressed
depended on the values of the final state masses). In particular, the
absence of the $1/m_Q$ corrections to the parton result for $m_{u,d}\sim
\beta\ll m_Q$ was very transparent. Additionally, the correspondence was
established between the step-by-step quark-gluon based OPE computations
and the matching contributions in the integrals determining the meson
formfactors in terms of the 't~Hooft wavefunctions. This allows one to
compare the OPE computations to the hadronic saturation at the
intermediate stages rather than only for the final result.

What could go wrong with the numerical analysis of Ref.\,\cite{gl}? We
note that the apparent ``effect'' was really small numerically and, in
the fiducial range of $m_Q/\beta \sim 8\; \mbox{to}\;10$
constituted only about
$1\%$. Moreover, a closer look at the plots of Ref.\,\cite{gl} shows
that the reported discrepancy somewhat increased toward the upper values
of $m_Q$. As a result, a better fit of the numerical points of
Ref.\,\cite{gl} would be achieved assuming small ${\cal O}(1/m_Q^0)$
(not power-suppressed!) corrections to the width, with the numerical
coefficient $\sim {\cal O}(10^{-2})$. Unfortunately, {\it a priori}
targeting the $1/m_Q$ terms, the authors did not explore the possibility
of alternative interpretations.

Incidentally, the scale of the claimed discrepancy lies just in the
magnitude range of the $1/m_Q^2$ effects from
the OPE which could be {\it a priori} expected at $m_Q \approx 10$.
It turns out that the numerical
points for $\Gamma(m_Q)$ in Ref.\,\cite{gl} can be well fitted by the
leading-$m_Q$ expression and adjustable
$\beta^2/m_Q^2$ terms with the coefficient ${\cal O}(1)$. Adding
arbitrary corrections $\sim \beta^3/m_Q^3$
with the larger coefficient up to $6\div 8$ allows
really perfect fits. Since the actual $1/m_Q^3$ corrections include the
expectation values of the four-quark operators which are not given by
factorization, they are uncertain. Their estimates showed that they
are typically
significantly enhanced compared to the naive dimensional estimate $\sim
\beta$.

Thus, the question would remain open before the actual OPE corrections are
computed. This was accomplished in  Ref.\,\cite{D2}.
While the radiative correction enhances the width by the factor
$1+\beta^2/(2m_Q^2)$ to order $1/m_Q^2$, the kinetic term
$-\mu_\pi^2/(2m_Q^2)$ tends to suppress it.\footnote{The value of
$\mu_\pi^2$ for quark masses used in Ref.\,\cite{gl} has been recently
evaluated by R.~Lebed to be about $0.8\beta^2$ (private communication).}
Therefore, if, as stated in
Ref.\,\cite{gl}, the plotted values refer to the bare mass $m_Q$, it
seems that the numerical calculation reported by Grinstein and Lebed
yielded the
result exceeding the asymptotic width by an amount ranging from a
fraction to a per cent.

The most apparent resolution, in our opinion, is that the analysis of
Ref.\,\cite{gl} simply does not control the accuracy of the numerical
computations of the amplitudes at the required level of a few per mill.
This is not surprising and roots to the well-known problems of numerical
computations. The width at large $m_Q$ is saturated by highly excited
states whose wavefunctions oscillate fast. This standard problem of the
numerical computations of the semiclassical transition overlaps is
additionally plagued by the general complexity of the numerical
solutions of the 't~Hooft equation and the encountered singular
integrals. These problems were alluded to in Ref.\,\cite{gl}.

The general lesson one draws from this comparison is not new: it is not
easy to correctly evaluate the actual asymptotic width in too
straightforward numerical summations of many exclusive widths, were it a
computer-simulated approximate solution for a theoretical model with --
theoretically speaking -- potentially unlimited accuracy, or the actual
experimental data.  The approximations made in practical applications
usually go across preserving the subtle interplay of different effects
which underlies delicate cancellations shaping the size and scaling of the
power suppressed nonperturbative effects.

The recent paper Ref.\,\cite{gl2} announced even more drastic numerical
mismatch of the actual hadronic width vs.\ the OPE considering WA, even
in the leading order at which the effect appears. In Sects.\,3 and 4, on
the contrary, we analytically computed the asymptotics of the
spectator-dependent widths and found exact correspondence
with the OPE. As a matter of fact, Ref.\,\cite{gl2} contains
self-contradicting assertions: It was stated that the (smeared)
current-current vacuum correlator (of the type determining, say,
$\sigma({\rm e^+e^-}\to\mbox{hadrons})$) is known to obey the OPE.
Simultaneously, the difference was claimed established for the effect of
WA in the context relying on the factorization expression Eq.\,(\ref{54}).
The starting expression -- whether correct or not for a particular model
-- thus equates the validity of the OPE for the WA nonleptonic width
to its applicability for the vacuum current correlator.
Fig.\,7 shows the ratio of the (smeared) values of the absorptive parts
of the correlator,
$$
\frac{{\rm Im}\,\Pi_{\mu\mu}(q^2)}{{\rm Im}\,\Pi_{\mu\mu}^{\rm tree}(q^2)}
$$
according to Ref.\,\cite{gl2}. The variance from unity, in general,
does seem to be present.

\thispagestyle{plain}
\begin{figure}[hhh]
 \begin{center}
 \mbox{\epsfig{file=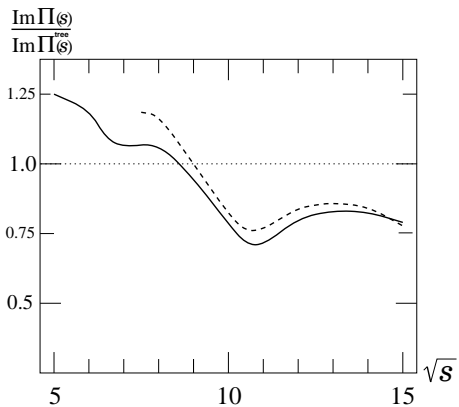,width=6cm}}
 \end{center}
 \caption{
The actual--to--partonic ratio of the absorptive part of the vector
current correlator at $m_u\!=\!m_d\!=\!0.56\beta$, according to
Ref.\,\protect\cite{gl2},
for $N_c=10$ (solid line) and $N_c=20$ (dashed line).
Energy scale is in units of $\beta$. Smearing procedure has been applied
for the hadronic cross section.
}
\end{figure}

We do not think that this warrants one to become cautious in applying
the OPE to the polarization operator. There are certain drawbacks in the
computations of Ref.\,\cite{gl2}. For unknown reasons
the authors presumed that the width Eq.\,(\ref{54})
becomes non-integrable for the large-$N_c$ theory and, therefore, cannot
be sensibly smeared.\footnote{It is obvious that for any regularization
consistent with unitarity, the width remains integrable, with the
integral around the spike independent of the regularization. This is
ensured by the dispersion relation; see Sect.\,3.}
As a result, instead of considering the current
correlator in the 't~Hooft model itself, a rather {\it ad hoc} ansatz was
adopted which was intended to mimic nonvanishing widths of the resonances
at finite $N_c$. The simple resonant representation for $\Pi(q^2)$,
\begin{equation}
\Pi(q^2)\;=\; N_c \sum_n \,\frac{c_n^2}{q^2-M_n^2+i\epsilon}\;,
\qquad
-{\rm Im}\, \Pi(q^2)\;=\;
\pi N_c \sum_n \,c_n^2\,\delta\left(q^2-M_n^2\right)
\label{160}
\end{equation}
was replaced by a complicated model where ${\rm Im}\, \Pi(q^2)$ was due
to the two-meson states, with their production amplitude containing the
resonance terms
$$
\frac{\sqrt{N_c} c_n}{q^2-M_n^2+ iM_n \Gamma_n}\;.
$$
At large $N_c$ the residues $\sqrt{N_c} c_n \sim \sqrt{N_c}$ while the
resonance decay
amplitudes ${\cal M}_{nkl} \sim 1/\sqrt{N_c}$; the widths $\Gamma_n \sim
1/N_c$.

As soon as $\Gamma_n$ are small enough, such a prescription does not
differ from the proper spectral density Eq.\,(\ref{160}) being a
concrete functional choice of the $\delta$-distributions,
\begin{equation}
\int_{|q^2-M_n^2|\gg M_n \Gamma_n} \;{\rm d}q^2\:
\frac{c_n^2}{|q^2-M_n^2+iM_n \Gamma_n|^2}\;=\; \pi\,c_n^2\,
\frac{1}{M_n \Gamma_n}\;,
\label{162}
\end{equation}
provided each $\Gamma_n$ is saturated by the included decay modes:
\begin{equation}
\sum_{k,l}|{\cal M}(n\,\to k\,l)|^2\cdot \Phi_{kl}\;=\;
2\,M_n\,\Gamma_n\;.
\label{163}
\end{equation}
Here $\Phi_{kl}$ denote the two-body decay phase space. Under these
constraints such a model -- if not true -- is at least legitimate.
However, two conditions must be observed: First, the resonances do
not overlap,
$\Gamma_n \ll |M_{n\pm 1}\!-\!M_n|$. Additionally, the partial decay
amplitudes and the phase space
factors must be practically constant within the (total) width of the
individual meson.

Both constraints are satisfied in the formal limit $N_c\to\infty$.
However, in practice the width grows with the mass of the resonance,
while the spacing between the successive ones decreases. Therefore, for
finite $N_c$ adopted in  Ref.\,\cite{gl2} (let alone the considered case
$N_c=1\,$!) the first condition was not well respected. Regarding the
second constraint, the problem gets additionally aggravated by the
singular two-body phase space in $D=2$. Therefore, it seems quite
probable that the reported disagreement is rooted to the inconsistencies
of the model for ${\rm Im}\, \Pi(q^2)$ adopted in Ref.\,\cite{gl2}.

A detailed look at the plots displayed there incidentally provides
support for this conjecture. The plots for (the smeared) ${\rm Im}\,
\Pi(q^2)$ show a rather unphysical shoulder at $m_Q\approx 10$ which
coincides with the mass of the resonance having abnormally large width
(Fig.\,9 of Ref.\,\cite{gl2}). On the other hand, the agreement between
the hadronic and quark widths is surprisingly good just at {\em lower}
masses where the resonances are more narrow. The point where the curves
start to diverge, apparently shifts upward with increasing $N_c$.

Finally, a simpler question remains open about the accuracy of numerical
computations employed in Ref.\,\cite{gl2}. Taken at face value, even
adopted
sampling of $m_Q$ when only 2 to 3 points fall inside a separate sharp
resonance
peak, seems insufficient to evaluate the smeared width reliably.
Whether these effects can explain the observed discrepancy $\sim 20\%$
at $m_Q \sim \;10\mbox{ to } 15$, remains to be
clarified.\footnote{Since the effect in question by itself is $1/q^2$ in
the polarization operator, it could be {\it a priori} conceivable to
have subleading $1/q^3$ corrections as large as $20\%$. However,
the OPE ensures that the corrections are suppressed by at least two
powers of $q$, and thus must be small; they are readily calculated.}

It is worth reiterating that the computational difficulties -- quite
significant in the analysis of Ref.\,\cite{gl2} -- to a large extent were
a hand-made
problem. Both $c_n$ and $M_n$ are readily computed without cumbersome
triple overlaps involving singular integrals, and in any case had been
determined to construct the authors' model. Computing the smeared width
directly from Eq.\,(\ref{160}) would be then quite straightforward.

\section{Conclusions}

We have examined the inclusive decay widths of heavy flavor mesons in
the 't~Hooft model in the context of the heavy quark expansion, paying
attention to the spectator-dependent effects sensitive to the flavor of
the spectator quark. To the order the high-energy
asymptotics
are calculated, there is no deviation from the OPE predictions, either for
semileptonic, nonleptonic decays or for the
${\rm e^+e^-}\to\mbox{hadrons}\,$-type processes -- as anticipated.

We confirm that there is no difference for the OPE whether a
semileptonic or nonleptonic width is considered. What matters is only
whether the particular observable can be represented as the {\it
complete} discontinuity of the properly constructed correlation
function over the cut in a suitable ``hard'' variable. This variable,
$\omega$, is the same for both semileptonic and nonleptonic  decays,
Eq.\,(\ref{72}). After that the only difference is the number of the
quark Green functions to be multiplied and to perform the expansion of
the product in the complex plane. The validity of the OPE, of course,
cannot depend on these technicalities.

This does not mean, however, that for all types of decays the
predictions of practical OPE truncated after the first few terms, work
with equal accuracy at a fixed mass $m_Q$. On the contrary, considerations
of Ref.\,\cite{five}
suggest that the
effective ``hardness'' scale can be smaller than literally the energy
release. Accordingly, higher onset of duality and larger deviations for
nonleptonic widths were obtained in Ref.\,\cite{inst} in the
instanton-based model.

Therefore, in our opinion, attempts to check (let alone to disprove)
the OPE itself in the concrete model are hardly meaningful beyond
illustrative purposes. What has a potential of providing useful insights,
is studying the behavior of the contributions violating local duality,
so far the least understood theoretically subject.

The ``practical'' OPE yields the width in the power expansion
\begin{equation}
\frac{\Gamma_{H_Q}}{\Gamma_0}\;=\;
A_0 + \frac{A_1}{m_Q} + \frac{A_2}{m_Q^2}\:+\;... \;\;.
\label{200}
\end{equation}
If the series in $1/m_Q$ were convergent (to the actual ratio),
$\Gamma_{H_Q}$ would have been an analytic function of $m_Q$
above a certain mass $m_0$ pointing to the onset of the {\em exact}
local parton-hadron duality. The actual $\Gamma_{H_Q}$ is definitely
non-analytic at any threshold (whether or not the amplitude vanishes at
the threshold). Thus, the `radius of convergence' cannot correspond to
the mass smaller than the threshold mass. Since in the actual QCD the
thresholds exist at arbitrary high energy, the power expansion in
Eq.\,(\ref{200}) can be only asymptotic, with formally zero radius of
convergence in $1/m_Q$.

In practice, the true threshold singularities are expected to be
strongly suppressed at large energies, and the corresponding
uncertainties in the OPE series quite small. In actual QCD they are
expected to be exponentially suppressed eventually, though, possibly,
starting at
larger energies. In the intermediate domain they can decrease as a
certain power and must {\em oscillate}.

As was illustrated in Ref.\,\cite{D2}, the power expansions like
Eq.\,(\ref{200}) are meaningful even beyond the power suppression where
the duality-violating oscillations show up. In the case of the heavy
quark widths where mass $m_Q$ cannot be varied in experiment, the size
of the duality-violating component may set the practical bound for
calculating the widths. Thus, it is important to have an idea about its
size. We emphasize that one should always include the leading QCD
effects to the partonic expressions, rather than compare the
actual observable with the bare quark result. In the model considered in
Ref.\,\cite{D2}, incorporating the power corrections from the practical
OPE suppressed the apparent deviations by more than an order of
magnitude.

It is worth noting that we identified the case
where the exact quark-hadron duality
in the interference width is saturated on a {\it single} final state. It is
realized in the chiral limit for the final state quarks. It seems to be
just an opposite case to the classical Small Velocity limit for
semileptonic decays noted by Shifman and Voloshin in 1986 \cite{volshif}.
While applicable only for large $N_c$ and in $D=2$, the
present case is peculiar in that the duality is not affected by
subleading power corrections. Both cases serve as a counter-example to
the lore often purportingly equating the accuracy of local duality to the
proliferation of the final state channels in the process in question.

The performed analysis elucidates how the general duality between the partonic
and hadronic widths works out its way at large energies. It can be
traced that, to the leading order in $1/m_Q$, the duality is simply the
completeness of eigenstates of the hadronic Hamiltonian. It is not even
required to explicitly solve the 't~Hooft equation to establish the
leading free quark result for the width, but just to know that the solutions
form a complete basis.\footnote{The Multhopp technique employed in
Ref.~\cite{gl} does not automatically respect orthogonality of solutions
with truncation and, therefore, in principle can lead to overestimating
even the leading-$m_Q$ coefficient.} The absence of the non-OPE terms already
at the next to leading order $1/m_Q$ is a dynamical fact requiring, for
example, a proper solution of the bound-state equations. This enters {\it
via} the average of the moments of the invariant hadronic mass in the
final state.

The 't~Hooft model gives an explicit example of the duality-violating
effects, at least of the ``minimal'', resonance-related nature. As a
general feature, we observe that\\
--- the duality between the appropriately averaged (to eliminate the
threshold singularities) hadronic widths and the truncated OPE predictions
sets in numerically well rather early, after only a couple of the principal
thresholds;\\
--- the local duality works much better in the heavy quark widths,
nonleptonic and semileptonic than in
$\sigma({\rm e^+e^-}\to\mbox{hadrons})$. This applies to both the
qualitative behavior (generally, a larger density of resonances
implying a more narrow minimal interval of smearing) and to the strength of
the threshold singularities as well as the threshold residues. The
qualitative discussion of the underlying reason can be found in
Ref.\,\cite{varenna}, Sect.\,3.5.3.

Can these lessons be transferred to actual QCD? Unfortunately, there are
some essential differences between it and the explored 't~Hooft model,
which must be important for the local duality violation.

The singular $1/|\vec{p}\,|$ two-body phase space in $D=2$ strongly
enhances the threshold singularities, compared to $|\vec{p}\,|$ in $D=4$.
In the actual QCD enhancement of the non-smooth 
behavior can rather be expected only from single resonances with 
masses close to $M_B$. While infinitely narrow at $N_c\to \infty$, 
they acquire significant width for $n_f/N_c$=1 which leads to drastic 
flattening of the resonance-related combs when the width becomes 
comparable to the distance between the successive resonances.  
Additionally, one expects a denser resonance structure, at least 
asymptotically, in the actual QCD in four dimensions (even at $N_c\to 
\infty$) than the equal in $M^2$ spacing in $D=2$. All this would 
lead to suppression of the duality violation.

Finally, similar types of the condensate corrections lead to a weaker
suppression in $D=2$ due to the smaller dimension of the corresponding
operators. This provides a larger room for various possible
nonperturbative effects.

These obvious differences would optimistically suggest that the
't~Hooft model represents, in a sense, an upper bound for violations of
local duality in QCD. While this is not excluded, such implications must
be regarded with caution.

In fact, the largest duality violating effects in the nonleptonic heavy
flavor decays can be expected from the resonance structures in the
combined $(\bar{q}q\bar{q}q)$ channel in the final state, embedding the
quarks belonging to both the ``semileptonic'' and ``hadronic''
subprocesses which do not decouple completely. Such states are lost in
the 't~Hooft model. Moreover, the limit $N_c\to \infty$ simply erases
the difference between the nonleptonic and semileptonic decays in this
respect.

Another peculiarity of QCD in two dimensions is absence of the real
gluonic degrees of freedom and of the perturbative logarithmic
short-distance corrections typical for $D\!=\!4$
(in contrast to the power-like ones in the
super-renormalizable $D\!=\!2$ QCD). So far we have no clues if the onset of
duality for excitation of the gluonic degrees of freedom is the same or
noticeably larger than for the processes describing the evolution of the
``valence'' quarks. Correspondingly, while observing a good qualitative
duality between the bare quark computations and the actual hadronic
probabilities already at energies $\sim 1\GeV$, one may have to ascend
to higher energies in the Minkowski domain to reach quantitative
agreement at the level of the
perturbative corrections. Let us note in this
respect that the `real gluons' in jet physics are reliably observed in
experiment only at rather high energies. Unfortunately, two-dimensional
theories do not allow one to study this interesting aspect of local
gluon-hadron duality even in the accessible model settings.

As emphasized above, we found that the OPE predictions universally hold
for the inclusive
widths. In this respect, the origin of the paradox in mismatch between
the size of the interference term on the quark and meson languages,
Eqs.\,(\ref{6}) and (\ref{8}), is quite transparent. The mismatch
emerges in the $N_c$-suppressed term. Addressing such effects requires
a more accurate control of decay amplitudes beyond the leading-$N_c$
counting rules. In particular, it is necessary to account for $1/N_c$
corrections in the color-allowed amplitudes rather than use
oversimplified prescription like Eq.\,(\ref{7}) (a possible resolution
was conjectured already in Ref.\,\cite{shifman}).

A closer look reveals that the problem originates just on the quark
configurations where all four quarks in the intermediate state have the
same color. This configuration is $1/N_c$ suppressed and
can be simply discarded for
the leading-$N_c$ probabilities. Within this color combination,
there is an ambiguity of allocating the quarks over the colorless
mesons which is absent otherwise. It is not difficult to see that the
naive computations based on the amplitude prescription Eq.\,(\ref{7})
amounts to counting the two possibilities independently, which is not
justified {\it a priori}. Moreover, it is a clear double counting at
least in the simple quark picture. To phrase it differently, the bases
of the final states used in the naive color rules like Eq.\,(\ref{7}),
are non-orthogonal beyond the leading order in $N_c$.

With the complicated confinement dynamics, we cannot know beforehand
what are the actual decay amplitudes to the particular hadronic states
in this case -- and, additionally, the simple two-meson picture of the
final state must be,
in general, extended when going beyond the leading order in $N_c$. However,
it seems
obvious that the naive prescription have little chance to be true in a
complete theory if it violates general requirements even in the simplest
case of almost free constituents.

This, incidentally, is an illustration of the fact that the naive
factorization of decay amplitudes {\it must} be violated at some point
at the $1/N_c$ level -- either in the corrections to the color-allowed
amplitudes, or at the leading level for color-suppressed amplitudes, or
in both. A dedicated study of the above inconsistency was undertaken in
the framework of the nonrelativistic quark model in Ref.\,\cite{pene}.
It was found that the problem is indeed resolved there in this way, with
the concrete dynamical mechanism dominating the modification of the
$1/N_c$-suppressed amplitudes depending on details of the model.

In Sect.\,4, considering the spectator-dependent preasymptotic correction
we formulated the problem of calculating the interference width
in the way that allowed us to avoid the complications associated with
the analysis of the $N_c$-subleading transition amplitudes, and to study
it in the theoretically clean environment. Incorporating the
spectator quark {\it via} PI made interference an effect
appearing in the same order in $N_c$ as the non-interference widths.
Then the  hadron-based computation could be performed consistently,
and the OPE prediction was readily reproduced.
\vspace*{.3cm}

{\bf Acknowledgments:} \hspace{.1em} N.U. would like to thank V.\ Petrov
for interesting discussions. We are deeply grateful to our collaborators
M.\ Shifman and A.\ Vainshtein for their generosity in sharing their
insights with us and for critical comments on the manuscript.
N.U. is grateful to M.~Burkardt for useful discussions and to R.~Lebed for a
number of clarifications and for
pointing out some omissions in the original text.
This work
was supported in part by the National Science Foundation under the grant
number PHY~96-05080 and by the RFFI under the grant number 96-15-96764.

\renewcommand{\thesection}{}
\section{Appendices}
\setcounter{section}{0}
\renewcommand{\thesubsection}{A\arabic{subsection}}

\subsection{'t Hooft wavefunctions in the semiclassical regime and the UV
divergences}

\renewcommand{\theequation}{A1.\arabic{equation}}
\setcounter{equation}{0}

Here we outline a more accurate determination of the scaling with
$\Lambda$ of the smearing parameter $\Delta$ in
Eqs.\,(\ref{86})-(\ref{91}).

At $M_n \gg \beta,\: m_{u,d}$ the solutions of the 't Hooft equations
are nearly free and approach the massless solutions \cite{thooft,stad}
\begin{equation}
\varphi_n(x) \;\simeq\; \sqrt{2} \cos{\pi n x}\;, \qquad
M_n^2 \,\approx\, \pi^2 \beta^2 n\;.
\label{A2}
\end{equation}
The first expression holds outside the end-point domains $x \to 0$ and
$x\to 1$ bounded by the ``classical turning points''
\begin{equation}
x,\;1-x\;\simeq \frac{m^2, \beta^2}{M_n^2}\;.
\label{A4}
\end{equation}
We again imply the simplified case of equal quark masses $m_u=m_d=m$.
The semiclassical wavefunctions (\ref{A2}) determine the
ultraviolet-singular part of the Green functions.

We shall adopt the exponential regularization of the sums. Then the
Green function $G(x,y;\Lambda)$ in Eq.\,(\ref{86}) has a meaning of the
Euclidean Green function for the 't~Hooft light-cone Hamiltonian, with
$1/\Lambda^2$ corresponding to the Euclidean (imaginary) light-cone
``time'' $ix_+$, and $x,y$ being ``coordinates'' in the space of the
light-cone momentum fractions. Formally, the problem is equivalent to
usual one-dimensional quantum mechanics on the interval $[0,1]$ with
Hamiltonian given by the r.h.s.\ of Eq.\,(\ref{30}). The first (local)
terms play a role of the potential whereas the integral term includes
the counterpart of the kinetic energy which is now approximately $\pi |k| $
for the ``momentum'' $k$ much larger than $\beta$.

The regularized summation is straightforward and yields literally
for $G(x,y;\Lambda)$
\begin{equation}
\frac{1}{4}\,\frac{\sinh{\epsilon}}{\sinh^2{\frac{\epsilon}{2}}
+ \sin^2{\frac{\pi}{2}(x-y)}}
\;+\;
\frac{1}{4}\,\frac{\sinh{\epsilon}}{\sinh^2{\frac{\epsilon}{2}}
+ \sin^2{\frac{\pi}{2}(x+y)}}
\;,
\qquad \epsilon\equiv \frac{\pi^2 \beta^2}{\Lambda^2}
\;.
\label{A7}
\end{equation}
Since the small-$n$ eigenfunctions can be quite different from
Eq.\,(\ref{A2}), there can be,
{\it a priori}, additional regular ($\epsilon$-independent)
terms in Eq.\,(\ref{A7}). However,
since at $\epsilon \to 0$ the limit of $G$ must yield $\delta$-function,
they are absent.

The first term in Eq.\,(\ref{A7}) clearly yields $\delta(x-y)$ at
$\epsilon \to 0$:
\begin{equation}
\frac{1}{4}\,\frac{\sinh{\epsilon}}{\sinh^2{\frac{\epsilon}{2}}
+ \sin^2{\frac{\pi}{2}(x-y)}}
\; \approx \;
\frac{\epsilon}{\epsilon^2+ \pi^2 (x-y)^2}
\;,
\label{A8}
\end{equation}
with the width $\Delta$ at finite $\epsilon$ amounting to
\begin{equation}
\Delta \;= \;
\frac{\epsilon}{\pi}\;=\; \frac{\pi\beta^2}{\Lambda^2}
\;.
\label{A9}
\end{equation}

The second term in Eq.\,(\ref{A7}) seems to appear when both $x$ and $y$
are small, $x,y\,\lsim \epsilon$ or $1-x, 1-y\,\lsim \epsilon$. In this
domain, however, wavefunctions $\varphi_n(x)$ with $n\lsim \Lambda^2$
are different (suppressed), and in reality this term must be discarded.
In practice the domain of such small $x$ must be treated separately,
which is illustrated below. Therefore, we adopt
\begin{equation}
G(x,y;\Lambda)\; \simeq\;
\frac{1}{4}\frac{\sinh{\epsilon}}{\sinh^2{\frac{\epsilon}{2}}
+ \sin^2{\frac{\pi}{2}(x-y)}}\;
\qquad \mbox{ at } \;
\frac{1}{\Lambda^2}\,\lsim x,y\,\lsim 1-\frac{1}{\Lambda^2} \;.
\label{A11}
\end{equation}

To address the end-point domains $x$ or $(1-x) \sim 1/\Lambda^2$ we need
to account for the corresponding behavior of $\varphi_n(x)$ in the
``classically forbidden''
domain, using the language of ordinary Quantum
Mechanics. In these domains the wavefunctions are suppressed:
\beq
\varphi_n(x) \; \sim\; \mbox{const }
\left\{
\begin{array}{ll}
\left(\frac{x}{x_t}\right)^{\gamma_1} \qquad \qquad & x\lsim x_t \ll
1
\\
\\
\left(\frac{1-x}{1-x_t}\right)^{\gamma_2} \qquad \qquad & 1-x\lsim 1-x_t \ll
1
\end{array}
\right.
\qquad x_t\approx \frac{m^2, \beta^2}{\Lambda^2}\;.
\label{A12}
\end{equation}

Now we consider the regularized sum rule Eq.\,(\ref{90}):
\begin{equation}
I_1(\Lambda^2) \;= \;
\int_0^{\infty} \,
q^2 {\rm d}q^2\: {\rm e}\,^{-q^2/\Lambda^2} \, \rho(q^2)\;=\;
\int_0^1 {\rm d}x\,{\rm d}y
\;\left(\frac{m^2}{x}+ \frac{m^2}{1-x}\right)\,G(x,y; \Lambda)\;.
\label{A13}
\end{equation}
Using $G(x,y; \Lambda)$ in the form Eq.\,(\ref{A11}) we get
\begin{equation}
I_1(\Lambda^2) \;= \;2\cdot m^2
\left[
\ln{\frac{\Lambda^2}{\pi\beta^2}} \;+\; \mbox{const}\,
\right]
\;,
\label{A15}
\end{equation}
where the factor $2$ reflects the contributions of both $x\to 0$ and
$x\to 1$. The logarithm is saturated at
\begin{equation}
|x-y|\lsim \frac{\beta^2}{\Lambda^2}\:, \qquad
\frac{\beta^2}{\Lambda^2} \ll x \ll 1
\label{A17}
\end{equation}
(and likewise with $x \to 1-x$, $\,y \to 1-y$). In this domain the
expression for $G(x,y; \Lambda)$ is legitimate.

Now we show directly that the domain $x\lsim \beta^2/\Lambda^2$ yields
only a constant, that is, a nonsingular contribution at $\Lambda \to
\infty$. First, we note that Eq.\,(\ref{83}) ensures an upper bound on
$c_n$, const$/M_n^2$.  More precisely,
\begin{equation}
|c_n| \;\le\; \mbox{const}\cdot \frac{m^2}{\gamma} \cdot
\frac{1}{M_n^2}\;
\propto \; \frac{1}{n}\;.
\label{A19}
\end{equation}
This, of course, follows also from
Eqs.\,(\ref{91}, \ref{94}).\footnote{While the
exact coefficient in Eqs.\,(\ref{90})-(\ref{94}) is a subtle thing
addressed here, the power of $n$ itself is simple enough.} Then we can
bound from above the small-$x$ contribution as
$$
\left|
\int_{x \lsim \frac{1}{\Lambda^2}}
{\rm d}x\,{\rm d}y
\,\left(\frac{m^2}{x}\!+\! \frac{m^2}{1\!-\!x}\right) G(x,y: \Lambda)
\right| \,
\le \, \mbox{const}\: \frac{m^2}{\gamma}
\sum_{n=1}^{\infty} \frac{{\rm e}\,^{-\frac{\pi^2 \beta^2 n}{\Lambda^2}}}{n}
\int_0^{\Lambda^{-2}} \!{\rm d}x
\:\frac{m^2}{x} \left(\pi^2 n x \right)^\gamma
$$
\begin{equation}
\le \; \mbox{const}\: \frac{m^4}{\gamma^2} \,
\sum_{n=1}^{\infty} \frac{{\rm e}\,^{-\pi^2 \beta^2 n/\Lambda^2}}{n}
\left(\frac{n}{\Lambda^2}\right)^{\gamma}
\;\simeq\; \mbox{const}\: \frac{m^4}{\gamma^3 \beta^2} \;.
\label{A22}
\end{equation}
Here we assumed the cutoff
in the integral over $x$ at $\sim 1/\Lambda^2$, up to an
arbitrary constant. The dimensionful factor can be either
$\beta^2$ or $m^2$, whichever is larger and determines the position of the
``classical turning'' point. The key property is the convergence of
both the integral
over small $x$ and the sum over $n$ due to the suppression of
$\varphi_n(x)$ in the ``classically forbidden''
domain.

Thus, the translation rule for the cutoff energy into the regularization
Eq.\,(\ref{87}) in the integral over $x,y$ is obtained directly from the
't~Hooft equation.

Using the similar technique, it is easy to establish the asymptotic
duality directly for the pseudoscalar current correlator
\beq
\Pi_{\!_{P}}(q^2) \;= \; N_c \, \sum_n \:\frac{d_n^2}{M_n^2-q^2}\;,
\qquad \;
\frac{1}{\pi} \, \Im \Pi_{\!_{P}}(q^2)\;= \; N_c \,\sum_n \:d_n^2\,
\delta(M_n^2\!-\!q^2) \;,
\label{A30}
\eeq
where
\bea
d_n\! = \sqrt{\frac{\pi}{N_c}} \; \langle n|\bar{d}
i\gamma_5 u |0
\rangle \!\!\!
&= &\; \qquad
\frac{1}{2}\,\int_0^1 {\rm d}x \left(\frac{m_d}{x}\!+\!
\frac{m_u}{1\!-\!x}\right)
\varphi_n(x)\\
\nonumber
&=&
\! \!\!\frac{1}{m_u\!+\!m_d}
\int_0^1 {\rm d}x \left(\frac{m_d^2}{x}\!+\!
\frac{m_u^2}{1\!-\!x}\right)
\varphi_n(x)\; \,
\mbox{ (even $n$; $0$ for odd $n$).}
\label{A32}
\eea
(The first form is a direct representation,
whereas the second expresses the
pseudoscalar density as the divergence of the axial current. A certain parity
relation \cite{callan} ascending to 't~Hooft
ensures they both hold.) To find the
asymptotics of
$d_n$ we consider the logarithmically divergent sum
$$
\int {\rm d}q^2 \,\frac{\rho_{\!_{P}}(q^2)}{q^2} \;\propto\;
\sum_n \, \frac{d_n^2}{M_n^2} \;=
$$
\beq
=\;
\sum_n \,
\frac{1}{2(m_u\!+\!m_d) M_n^2}\,
\int_0^1 \!{\rm d}x
\left(\frac{m_d}{x}\!+\!\frac{m_u}{1\!-\!x}\right)
\varphi_n(x)\:
\int_0^1 \!{\rm d}y
\left(\frac{m_d^2}{y}\!+\!\frac{m_u^2}{1\!-\!y}\right)
\varphi_n(y)
\:.
\label{A35}
\end{equation}
Eq.\,(\ref{83}) allows one to rewrite it as
\beq
\sum_n \frac{d_n^2}{M_n^2} =
\frac{1}{2(m_u\!+\!m_d)}\,
\sum_n \,
\int_0^1 \!{\rm d}x
\left(\frac{m_d}{x}\!+\!\frac{m_u}{1\!-\!x}\right)
\varphi_n(x)\,
\int_0^1 \!{\rm d}y \,\varphi_n(y)
\;,
\label{A37}
\end{equation}
which returns us to Eq.\,(\ref{84}).

\subsection{$1/m_Q$ expansion of the heavy quark weak decay amplitudes}

\renewcommand{\theequation}{A2.\arabic{equation}}
\setcounter{equation}{0}

As discussed in Sect.\,5, the perturbative corrections to the weak decay
vertex appear to order $\beta^2$. In the light-front formalism they
vanish at the kinematic point $q^2=0$ \cite{D2}, since it can be
realized as the configuration with $q_-=0$, for which no physical states
in the $t$ channel is possible. The loop corrections proportional to
the dimensionful coupling $\beta^2$ can be, therefore, inversely
proportional to $m_Q^2$ or $E_{\rm rel}^2 \sim (m_Q-m_q-\sqrt{q^2})^2$.
Since at
$m_Q-m_q-\sqrt{q^2} \sim \beta$ the process is `soft', we do not
consider here this domain, and do not distinguish between the scales of
$m_Q$ and $E_{\rm rel}$. It is important that no dynamic gluon degrees of
freedom exist in $D=2$. Therefore, in the 't~Hooft model vertex
corrections separately are perturbatively infrared finite in physical
gauges. This was explicitly illustrated in Ref.\,\cite{D2}.

As a result, the actual transition amplitudes $\langle k| J |H_Q\rangle$
up to terms $\beta^2/m_Q^2$ must be given by only some overlaps of the
initial and final state wavefunctions, at arbitrary (though not too close
to $(m_Q-m_q)^2$) values of $q^2$.
We assume that the initial state $H_Q$ is either
a ground state or has a finite (not scaling with $m_Q$) excitation
number. The final state $k$ can be arbitrary. We will show that the
explicit expressions for the transition amplitudes in the 't Hooft model
exhibit this parton-deduced property.

Let us consider, for example, the representation used in Ref.\,\cite{gl},
although we put it in a slightly modified form similar to that of
Refs.\,\cite{burkphd,burkswan}. One introduces the
kinematic variable $\omega$ which
depends on $q^2$ and the final state meson mass:
\beq
\omega\;=\;\frac{1}{2}\,\left[1\;+\;\frac{q^2-M_k^2}{M_{H_Q}^2}\;-\;
\sqrt{1-2\frac{q^2+M_k^2}{M_{H_Q}^2} +
\left(\frac{q^2-M_k^2}{M_{H_Q}^2}\right)^2}\;
\right]\;,
\label{B4}
\eeq
which, in the light-cone formalism
has a meaning of the fraction of the momentum seen in the
infinite-momentum frame carried by the particle with mass $\sqrt{q^2}$
in the two body decay of $B$ meson, if another particle
has mass $M_k$. This fraction has two possible values corresponding to
the two possible directions of meson $k$ in the rest frame.
We chose the above branch to have $q_-\to 0$ as $q^2\to 0$, as in
Ref.\,\cite{D2}.
The light-cone fraction $\omega$ has a very simple meaning in the rest
frame as well:
\beq
(1\!-\!\omega) M_{H_Q}\;=\; (|\vec{p}_k|+E_k)_{\rm c.m.}\;,
\qquad \; \omega M_{H_Q} = (-|\vec{q}\,|+q_0)_{\rm c.m.}
\;.
\label{B4a}
\eeq
Two $\omega$-dependent $H_Q\,D^{(k)}$ overlaps are then
considered,
$$
{\cal C}_{k}(\omega)\;=\;-\frac{1\!-\!\omega}{\omega}\,\int_0^1\, {\rm d}y\;
\varphi_k(y) \varphi_{H_Q}(\,1-(1\!-\!\omega)(1-y)\,)\;,
$$
\beq
{\cal D}_{k}(\omega)\;=\;-\omega\,\int_0^1\, {\rm d}y\;
\frac{\varphi_k(y)}{y}
\frac{\varphi_{H_Q}(\,1-(1\!-\!\omega)(1-y)\,)}{1-(1\!-\!\omega)(1-y)}
\label{B5}
\eeq
which correspond to the effects unrelated to vertex corrections.

Additionally, in the $t$ channel of the decay process one can have
various bound states of
$B_c$ mesons; we reserve index $l$ for them, and their masses will be
denoted $\mu_l$. For each $t$-channel resonance there is an
$\omega$-dependent triple-meson (``$H_Q B_c^{(l)}D^{(k)}$'') overlap
\beq
F_{lk}(\omega) = \omega(1\!-\!\omega) \int_0^1 \! {\rm d}x \int_0^1 \!
{\rm d}y
\frac{\varphi_l^{B_c}(x)\varphi_k(y)}{[\omega(1\!-\!x)\!+
\!(1\!-\!\omega)y]^2}\:
\left\{ \varphi_{H_{\!Q}}(\omega x)\!-\!
\varphi_{H_{\!Q}}(1\!-\!(1\!-\!\omega)(1\!-\!y))
\right\} \!.
\label{B7}
\eeq
With these notations one has for the transition amplitudes
\beq
\nonumber
\epsilon_{\mu\nu} q_\mu \,\langle k|\,\bar{q}\gamma_\nu Q\, |H_Q(P)\rangle
=
\frac{\beta^2 \sqrt{\pi}}{\sqrt{N_c}} \sum_l
\frac{q^2+(-1)^l \mu_l^2}{q^2-\mu_l^2}\,
f_l F_{lk}(\omega)\:-\, q^2 {\cal C}_{k}(\omega)\,+\,
m_Q m_q {\cal D}_{k}(\omega)
\, ,
\eeq
\beq
\nonumber
q_\mu \,\langle k|\,\bar{q}\gamma_\mu Q\,
|H_Q(P)\rangle
=
\frac{\beta^2 \sqrt{\pi}}{\sqrt{N_c}}
\sum_l \frac{-q^2 \!+
\! (\!-1\!)^l
\mu_l^2}{q^2-\mu_l^2}\,
f_l F_{lk}(\omega)\:+\, q^2 {\cal C}_{k}(\omega)\,+\,
m_Q m_q {\cal D}_{k}(\omega)
\,.
\label{B8}
\eeq
Here, for convenience, we wrote the invariant combinations of the
amplitudes instead of the two Lorentz components of the current
separately.
For example, the decay amplitude $B\ra
D^{(k)}+\pi^{(n)}$ in the considered case of vector-like interactions
takes the form
\beq
{\cal M}_{kn}\;=\; \frac{G}{\sqrt{2\pi}}\: c_n\,\left[
\sum_l \frac{-P_n q^2+(-1)^l \mu_l^2}{q^2-\mu_l^2}\,
c_l F_{lk}(\omega)\;+\; P_n q^2 {\cal C}_{k}(\omega)\;+\;
m_Q m_q {\cal D}_{k}(\omega)\:
\right]\;,
\label{B9}
\eeq
where $P_n$ is the parity $(-1)^{n+1}$ of the $\pi^{(n)}$ state. Here
and below throughout this Appendix we suppressed the factors $N_c$ and put
dimensionful $\beta=1$.
It is easy to see that at $q^2=0$ ($\omega \to 0$) only the term $\sim
{\cal C}_{k}$ survives and this expression indeed reduces to
Eqs.\,(\ref{105}, \ref{110}).

We shall show that the first term in this relation associated with the
dynamics in the $t$ channel,  is
$1/m_Q^2$-suppressed  compared to the terms given by overlaps
${\cal C}$ and ${\cal D}$. This implements the approximate ``on-shell''
condition
for the heavy quark $Q$; in this case it can be accomplished employing
the $1/m_Q$ expansion of the initial-state wavefunction $\varphi_{H_Q}$.

\subsubsection{Nonrelativistic expansion in the 't Hooft equation}

To analyze the heavy quark system, it is advantageous to introduce the
nonrelativistic variables:\footnote{This standard for the infinite momentum
frame procedure in the context of the 't~Hooft model was first
considered in Refs.\,\cite{burk,burkswan}.}
\beq
M_n\;=\;m_Q\,+\,\epsilon_n\;,\qquad t\,=\,(1-x)m_Q \qquad
\mbox{and}\qquad \Psi_n(t)\,=\,\frac{1}{\sqrt{m_Q}}\:
\varphi_n\left(1-\frac{t}{m_Q}\right)\;,
\label{B75}
\eeq
in terms of which the equation takes the form
\beq
\left(\epsilon_n \!+ \!\frac{\epsilon_n^2\!+\!1}{2m_Q}\right)
\Psi_n(t) \;=\;
\left(\frac{m_\sp^2\!-\!1}{2t} + \frac{t}{2}
\frac{1\!-\!\frac{1}{m_Q^2}}{1\!-\!\frac{t}{m_Q}}
\right) \Psi_n(t)\;-\;
\frac{1}{2}\, \int_{0}^{m_Q} {\rm d}s\:\frac{\Psi_n(s)}{(t-s)^2}\;\;.
\label{B76}
\eeq
The asymptotics of $\Psi_n(t)$ at $1\ll t \ll m_Q$ is given by
\beq
\Psi_n(t) \;\simeq \; \frac{F_n}{t^3}\qquad
\mbox{with} \qquad F_n\;=\;\sqrt{\pi m_Q}\,f_n\;=\;\int_0^{m_Q}\,
dt\; \Psi_n(t)\;\;.
\label{B77}
\eeq
One can extend the bound-state problem Eq.\,(\ref{B76}) from $[0,m_Q]$ to
the
whole interval $[0,\infty)$ if the exact linear potential term
$\frac{t}{2}\left((1-1/m_Q^2)/(1-t/m_Q)\right)$ is replaced by its
formal expansion in $t/m_Q$. This is justified literally only up to
$1/m_Q^4$ terms ($1/m_Q^3$ in the wavefunction), when the ultraviolet
divergences of the static heavy quark
expansion first show up.
Further terms can be obtained using the explicit large-$t$ asymptotics
in Eq.\,(\ref{B77}) and the end-point behavior of $\Psi_n(t)$ at $t\to
m_Q$. For example, the improved nonrelativistic equation takes the form
$$
\left(\epsilon_n \!+ \!\frac{\epsilon_n^2 \!+ \!1}{2m_Q}\right)
\Psi_n(t) \,=\,
\frac{m_\sp^2 \! -\! 1}{2t}\, \Psi_n(t) + \frac{1}{2}
\left[\left(1\!-\!\frac{1}{m_Q^2}\right) t +
\left(1\!-\!\frac{1}{m_Q^2}\right) \frac{t^2}{m_Q} + \frac{t^3}{m_Q^2}
\right]\, \Psi_n(t)
$$
\beq
-\;
\frac{1}{2}\, \int_{0}^{\infty}\; {\rm d}s\,
\Psi_n(s)\,\frac{1}{(t-s)^2}
\;\;.
\label{B79}
\eeq
Eq.~(\ref{B79}) can be viewed as a usual variational problem for the
non-local
Hamiltonian defined as
$$
\matel{\Psi}{{\cal H}}{\Psi'}\,=\,
\int_{0}^{\infty} \! {\rm d}t\, \Psi(t)
\left[
\frac{m_\sp^2\!-\!1}{2t} + \frac{1}{2}
\left(1\!-\!\frac{1}{m_Q^2}\right) t +
\left(1\!-\!\frac{1}{m_Q^2}\right)\frac{t^2}{2m_Q} +
\frac{t^3}{2m_Q^2}
\right]\, \Psi'(t)
$$
\beq
-\;
\frac{1}{2}\,
\int_{0}^{\infty}\; {\rm d}t\, \int_{0}^{\infty}\; ds
\frac{\Psi(t)\Psi'(s)}{(s-t)^2}
\;\;.
\label{B80}
\eeq
For example, the analogue of the nonrelativistic expansion for the heavy
hadron mass takes the form
\beq
M_{H_Q}\!-\!m_Q\,=\, \aver{t} - \frac{1}{2m_Q} +
\frac{3\aver{t^2}\!-\! \aver{t}^2}{2m_Q}  +
\frac{4\aver{t^3} \!-\! 3\aver{t}\aver{t^2}\!+\!\aver{t}^3 }{2m_Q^2}-
\frac{\aver{t}}{2m_Q^2}\,+\,{\cal O}\!\left(\frac{1}{m_Q^3}\right)
\,,
\label{B84}
\eeq
with the average $\aver{...}$ defined in the standard way as the
integral over $t$ with the weight $|\Psi(t)|^2$. Here all averages are
calculated with the finite-$m_Q$ $\Psi(t)$.
To exclude the Coulomb interaction term we used certain relations which
are derived in the way analogous to
the virial theorem in Quantum Mechanics. Namely, for any eigenfunction
$\Psi_n(t)$ we can consider
the average of ${\cal H}$ over the trial function
$\sqrt{\lambda}\,\Psi_n(\lambda t)$, and require a minimum at
$\lambda=1$ (a similar trick was used in Ref.\,\cite{burk} for the case of
infinite $m_Q$).
In this way one obtains
\beq
(m_\sp^2-1)\aver{\frac{1}{t}} \;-\; \aver{{\cal V}} \;-\;
\left(1-\frac{1}{m_Q^2}\right)
\aver{t}\;-\; 2\,\frac{\aver{t^2}}{m_Q}\;-\; 3\,\frac{\aver{t^3}}{m_Q^2}
\;= \;{\cal O}\left(\frac{1}{m_Q^3}\right)
\label{B82}
\eeq
while
\beq
\epsilon_n + \frac{\epsilon_n^2+1}{2m_Q} =
\frac{1}{2}\left((m_\sp^2-1) \aver{\frac{1}{t}}-\aver{{\cal V}}
\right)+
\frac{1}{2}\left(1-\frac{1}{m_Q^2}\right)
\aver{t}+  \frac{1}{2}\frac{\aver{t^2}}{m_Q}+
\frac{1}{2}\frac{\aver{t^3}}{m_Q^2}
\;+\;{\cal O}\left(\frac{1}{m_Q^3}\right),
\label{B83}
\eeq
where $\aver{{\cal V}}$ denotes the expectation value of the
integral term in the equation.

The bound-state--independent term $-1/2m_Q$ in Eq.\,(\ref{B84})
deserves a special note: it
represents the short-distance renormalization of the bare mass we used,
originating from momenta $\sim m_Q$. There is no infrared part of
the mass:  it enters at the scale much larger than $\Lam$. The effect of
smaller momenta is described by the nonperturbative 't~Hooft
wavefunction rather than infrared-divergent perturbative diagram.

Let us note that the operator $t/2$ is associated with the breaking
of scale-invariance of the static 't~Hooft equation ({\it i.e.}
$\Psi(\lambda t)$ is the solution of the equation with the linear term
$\lambda^2 t/2$).
In other words, the commutation relation holds
$\Big[t\frac{\rm d}{{\rm d} t}, {\cal H}_{\infty}\Big] \!= \!
t\!-\!{\cal H}_{\infty}$, where ${\cal H}_{\infty}$ is the static
't Hooft Hamiltonian, Eq.\,(\ref{B80}) with $m_Q
\!=\!
\infty$. This fact can be used to obtain the
variation of
$\aver{t}$ when including perturbations, in the analogy to the case of
QCD (Ref.\,\cite{optical}, Sect.\,II):
$$
\delta\, \aver{t}\; = \; -l \,\aver{\delta{\cal H}_l}
$$
for the perturbation $\delta{\cal H}_l$ which is a homogeneous rank-$l$
functional of $t$. This follows from the usual operator relations like
$\langle iT\{[{\cal H},A],B\}\rangle =\langle [A,B]\rangle $.
For instance, to order $1/m_Q$
one has $\aver{t}=\aver{t}_\infty- \aver{t^2}/m_Q$, so that in
Eq.\,(\ref{B84}) $M_{H_Q}-m_Q \simeq \aver{t}_\infty +
\aver{t^2-\bar{t}^{\,2}}/(2m_Q)$, as it should be.

\subsubsection{Nonrelativistic expansion of the decay amplitude}

In the $1/m_Q$ expansion of the $H_Q\to D^{(k)}$ amplitude we pass from
$\varphi_{H_Q}(x)$ to the nonrelativistic wavefunctions $\Psi_{H_Q}(t)$
{\it and}, in the case of the triple vertices $F_{lk}$, rewrite
$\varphi_l^{B_c}(x)$ {\it via} the corresponding $\Psi_l^{B_c}(u)$. The
nonrelativistic approximation for $c$ quark is not employed here,
however. Then
\beq
{\cal C}_{k}(\omega)\;=\;-\frac{1}{\omega\sqrt{m_Q}}\,
\int_0^{(1\!-\!\omega)m_Q}\, {\rm d}t\;\Psi_{H_Q}(t)\,
\varphi_k\left( 1-\frac{t}{(1\!-\!\omega)m_Q}\right)
\;,
\label{153}
\eeq
\beq
{\cal D}_{m}(\omega)\;=\;-\frac{\omega}{(1\!-\!\omega)\sqrt{m_Q}}\,
\int_0^{(1\!-\!\omega)m_Q}\, \frac{dt}{1-t/m_Q}\;\Psi_{H_Q}(t)\,
\frac{\varphi_k\left( 1-\frac{t}{(1\!-\!\omega)m_Q}\right)}
{1-\frac{t}{(1\!-\!\omega)m_Q}}
\;,
\label{154}
\eeq
and, for the triple vertex,
$$
F_{lk}(\omega)\;=\; \omega(1\!-\!\omega)\; \times
$$
\beq
\! \int_0^{m_Q} \! {\rm d}u\,\int_0^1 \! {\rm d}y
\frac{\Psi_l^{B_c}(u)\varphi_k(y)}{\left[(1\!-\!\omega)y\!+\!
\frac{\omega u}{m_Q}\right]^2}
\left\{ \Psi_{H_Q}(\omega u + (1\!-\!\omega) m_Q) \!-\!
\Psi_{H_Q}( (1\!-\!\omega)(1\!-\!y) m_Q) \right\}
.
\label{152}
\eeq
At arbitrary $q^2$ we have $\omega\sim 1$; nevertheless, at large enough
energy release $(1\!-\!\omega)m_Q \gg 1$ still holds. This parameter defines
the `hardness' and is used
in the $1/m_Q$ expansion.

The nonrelativistic expansion amounts to assuming that the
support of the $\Psi$ functions is limited to a finite interval
of the argument of order $1$, and extending the integration over $t$
and $u$ to infinity.
With the fiducial domain of integration $t,u \sim 1$,
it is readily seen
that ${\cal C}_{k}$ and ${\cal D}_{k}$ scale like $m_Q^{-1/2}$ and lead
to the properly $m_Q$-behaved transition amplitudes. On the other hand,
$F_{lk} \sim 1/m_Q$ and are accompanied by the factors $c_l\sim
m_Q^{-1/2}$. The sum over $l$ is effectively cut off above $l \sim m_Q$
where $\mu_l$ exceeds $m_Q$. Altogether, the terms with $F_{lk}$ yield
corrections to the decay amplitudes suppressed by at least $1/m_Q^2$.

The leading heavy-quark transition amplitudes governed by the overlap
factors ${\cal C}$ and ${\cal D}$, Eqs.\,(\ref{153})--(\ref{154}),
exhibit explicitly the proper functional dependence on the combination
of $m_Q$ and $q^2$: the inner product of wavefunctions depends only on
$(1\!-\!\omega) m_Q$ whose value just fixes the energy (or momentum)
of the final state hadron in the rest frame, Eq.\,(\ref{B4a}).
In reality, this property of the leading-$m_Q$ transition amplitudes is more
general and is not related to the smallness of the perturbative
corrections. It holds at small energy release as well, where the vertex
corrections are not power suppressed; this is addressed below in the end of
Appendix~2.

\subsubsection{``Semileptonic'' width $\Gamma_{\rm sl}(q^2)$ at
arbitrary $q^2$ with $1/m_Q$ accuracy}

Here we illustrate how the parton expression for $\Gamma_{\rm sl}(q^2)$
is reproduced at nonzero $q^2$. We assume that the width averaged over
an interval of $m_Q$ or $q^2$ is considered, so that the threshold
factors become nonsingular. The analysis differs from the case of
$q^2=0$ anatomised in Ref.\,\cite{D2} only in technical details. In
particular, the width is still saturated by the states with $M_k^2\lsim
(1\!-\!q^2/m_Q^2)m_Q$, and the summation over the final states can be extended
to infinity. The decay amplitudes are given by the wavefunction overlaps
to this accuracy.
Let us assume for simplicity that $m_c \ll m_Q$. Then only the term
$\sim {\cal C}_{k}$ survives:
$$
\Gamma(q^2)\;=\; \frac{G^2}{4\pi}\:
\sum_k\, \frac{1}{2M_{H_Q}^2 |\vec{p}_k|}\,
\frac{q^4}{\omega_k^2 m_Q}\; \times
$$
\beq
\int_0^{(1\!-\!\omega_k)m_Q} {\rm d}t\:
\int_0^{(1\!-\!\omega_k)m_Q} {\rm d}s\; \Psi_{\!H_{\!Q}}(t)\,
\Psi_{\!H_{\!Q}}(s)\;
\varphi_k\left(\!1-\!\frac{t}{(1\!-\!\omega_k)m_Q}\right)\,
\varphi_k\left(1\!-\!\frac{s}{(1\!-\!\omega_k)m_Q}\right)
\,.
\label{B165}
\eeq
The leading term in the expansion of the width emerges
if we neglect $M_k$ compared to $(1\!-\!\omega)m_Q$. Then
$\omega_k=q^2 / M_{H_Q}^2$ and $2|\vec{p}_k|= (1\!-\!\omega_k)M_{H_Q}$
are $k$-independent, and we get up to the power corrections
\beq
\Gamma(q^2) =  \frac{G^2}{4\pi}\: \frac{1}{1\!-\!\omega}
\int_0^{\infty} \!{\rm d}t \,{\rm d}s\; \Psi_{\!H_{\!Q}}(t)\,
\Psi_{\!H_{\!Q}}(s)\;
\delta\left(\frac{t-s}{(1\!-\!\omega)m_Q}\right)
 =
\frac{G^2 m_Q}{4\pi}\:\left[1\,+\,
{\cal O}\!\left(\frac{1}{m_Q}\right)\right]\,.
\label{B167}
\eeq
Account for the $1/m_Q$ effects requires expanding $|\vec{p}_k|$ and
$\omega_k$ in $M_k^2/(m_Q^2-q^2)$, and leads to the `second' sum
rule for the average of $M_k^2$ \cite{D2}. Here is how it works.

The partial width $\Gamma_k(q^2)$ to $1/m_Q$ accuracy  takes the form
$$
\Gamma_k(q^2)\;=\; \frac{G^2}{4\pi}\:
\frac{M_{H_Q}^2}{m_Q (1\!-\!\omega_0)}
\left\{
(1\!-\!\omega_0)m_Q \;+ \;\frac{M_k^2}{m_Q^2} \,\times
\right.
$$
$$
\left. \left[-\frac{1}{(1\!-\!\omega_0)}
\int {\rm d}t \,{\rm d}s \:
\Psi_{H_Q}(t)\, \Psi_{H_Q}(s)\;
\varphi_k \!\left(1-\frac{t}{(1\!-\!\omega_0)m_Q}\right)\,
\varphi_k \!\left(1-\frac{s}{(1\!-\!\omega_0)m_Q}\right)
\,+
\right.\right.
$$
\beq
\left.\left.
+\,
\frac{2\omega_0}{(1\!-\!\omega_0)^2}
\int  {\rm d}t \,{\rm d}s \;
t\,\Psi'_{\!H_{\!Q}}(t)\, \Psi_{\!H_{\!Q}}(s)\;
\varphi_k \!\left(1-\frac{t}{(1\!-\!\omega_0)m_Q}\right)\,
\varphi_k\! \left(1-\frac{s}{(1\!-\!\omega_0)m_Q}\right)
\right]
\right\}
\,,
\label{B169}
\eeq
where $\omega_0=q^2/M_{H_Q}^2$.

Using the relation
\beq
\sum_k\: M_k^2 \; \varphi_k(x)\,\varphi_k(y)\;=\;
\left( \frac{m_c^2-1}{x}+\frac{m_\sp^2-1}{1-x} \right)\, \delta(x-y) \;-\;
\frac{1}{(x-y)^2}
\label{B171}
\eeq
and equation (\ref{B79}) to the leading order in $1/m_Q$, we
obtain
$$
\sum_k\: M_k^2 \;
\int\, {\rm d}s\;\Psi_{H_Q}(s)\;
\varphi_k\left(1-\frac{t}{(1\!-\!\omega_0)m_Q}\right)\,
\varphi_k\left(1-\frac{s}{(1\!-\!\omega_0)m_Q}\right)
\;=
$$
\beq
=\;
(1\!-\!\omega_0)^2 m_Q^2\left( 2\epsilon -t \right)\; \Psi_{H_Q}(t)
\label{B173}
\eeq
($\epsilon=M_{H_Q}-m_Q$). Hence, we arrive at
$$
\Gamma(q^2)= \frac{G^2}{4\pi}\:
M_{H_Q} \left\{
1+ \frac{1}{m_Q} \left[
\int {\rm d}t\, (t-2\epsilon)\,\Psi_{H_Q}^2(t) \;+
\right.\right.
$$
\beq
\left.\left.
+\;
\frac{2\omega_0}{1\!-\!\omega_0}\,
\int {\rm d}t\, t(2\epsilon-t)\:\Psi'_{H_Q}(t)\,\Psi_{H_Q}(t)\;
\right]
+ {\cal O}\left(\frac{1}{m_Q^2}\right)\,
\right\}\,
.
\label{B175}
\eeq
To evaluate the last integral we note that for any $f(t)$
\beq
\int {\rm d}t\; \Psi'_{H_Q}(t)\,\Psi_{H_Q}(t)\:f(t)\:=\:
\frac{1}{2} \int {\rm d}t\,
\left[
\frac{{\rm d}}{{\rm d}t} \left( \Psi_{H_Q}^2(t)\:f(t) \right)
\,-\,
\Psi_{H_Q}^2(t) \:f'(t)
\right]
\,=\:
-\frac{1}{2}\,\aver{f'(t)}
\:.
\label{B177}
\eeq
Therefore, Eq.~(\ref{B175}) takes the form
\beq
\Gamma(q^2)= \frac{G^2 M_{H_Q}}{4\pi}\: \left\{
1\!-\!\frac{2\epsilon\!-\!\aver{t}}{m_Q}
\!+\!
\frac{4\omega_0}{1\!-\!\omega_0}\,\frac{\aver{t}\!-\!\epsilon}{m_Q}
+{\cal O}\!\left(\frac{1}{m_Q^2}\right)
\right\}
=
\frac{G^2 m_Q}{4\pi}\:
\left\{
1+{\cal O}\!\left(\frac{1}{m_Q^2}\right)
\right\}
\!,
\label{B178}
\eeq
where we recalled that $\epsilon-\aver{t}=
{\cal O}\left(\frac{1}{m_Q}\right)\;$.

The next, $1/m_Q^2$ corrections, on the other hand, are not only
limited to the term
$-\frac{\mu_\pi^2}{2m_Q^2}$, but include also those coming from the
perturbative
vertex renormalization and from the sensitivity of the decay width
$Q\to q + \phi$ with $m_\phi^2 \ne 0$
to the short-distance quark mass renormalization $m^2 \to m^2-\beta^2$.

\subsubsection{Decays at maximal $q^2$}

At $q^2$ close to the energy release, $(m_Q-m_q)^2 -q^2 \lsim \beta m_Q$
(or at $(m_Q-m_q) \lsim \beta\,$) the decay processes are not hard but
proceed over the time intervals $\gsim 1/\beta$. As such, they in
general are sensitive to the $t$-channel evolution as well. OPE does
not allow to compute these widths in the short-distance expansion
directly even to the leading order in $1/m_Q$. However, it relates the
overall width associated with this domain of $q^2$ ({\it i.e.},
integrated over $q^2$) to the expectation value of the local
four-fermion operator:
\begin{equation}
\Gamma^{\rm end\,point}\;\simeq\;G^2\:
\bar{\rho}(m_Q^2) \cdot
\left(P_\mu P_\nu - \delta_{\mu\nu} M_{H_Q}^2 \right)
\frac{1}{2M_{H_Q}}
\langle H_Q(P)|\,\bar{Q}\gamma_\mu q\, \bar{q}\gamma_\nu Q\, |H_Q(P)\rangle
\label{B70}
\end{equation}
(this fact was proven in \cite{WA} although was used to calculate
certain preasymptotic
corrections to the heavy quark widths since the original papers
\cite{vsold}). Complicated strong interaction dynamics shows up here as
a nontrivial expectation value of this operator (for light $q$; it
is perturbatively calculable when $q$ is heavy). In the 't~Hooft model
it essentially depends on the details of the lowest $t$-channel states. As
has been already discussed a few times, a smeared decay width is assumed
here, which is reflected in Eq.\,(\ref{B70}): it incorporates the spectral
density $\bar{\rho}(m_Q^2)$ averaged over an interval. With a
continuous $\rho(q^2)$ there would be no resonance structure in
$\Gamma(m_Q)$.

Referring to paper \cite{WA} for the formal derivation of
Eq.\,(\ref{B70}), here we illustrate it schematically in a transparent way.
Let us consider a particular meson $k$ in the final state. The
corresponding partial decay probability (in the rest frame) is given by
$$
\Gamma_k(q^2) \;=\;
\frac{G^2}{2}\:\frac{2}{8M_{H_Q}^2 |\vec{p}_k(q^2)|}\: \frac{1}{\pi}
\rho(q^2)
\left(q_\mu q_\nu - \delta_{\mu\nu} q^2 \right)\;
\times
$$
\begin{equation}
\langle H_Q(P)|\,\bar{Q}\gamma_\mu q(0)\,|k(\vec p_k)\rangle
\langle k(\vec p_k)|\, \bar{q}\gamma_\nu Q(0)\, |H_Q(P)\rangle
\;.
\label{B72}
\end{equation}
Integrating over some interval of $q^2$ we can pass to the variable
$|\vec p_k|$ according to
$$
q_0=\frac{M_{H_Q}^2+q^2-M_k^2}{2M_{H_Q}}\,, \qquad
{\rm d}q_0 = -{\rm d}E_k\, \qquad
E_k \,{\rm d}E_k = |\vec{p}_k|\, {\rm d}|\vec{p}_k|\;,
$$
and have
$$
\int {\rm d}q^2\:\Gamma_k(q^2)\;=\;
G^2\, \int \frac{2\,{\rm d}|\vec{p}_k|}{2\pi\, 2E_k}
\:\rho(q^2(|\vec{p}_k|)) \left((P-p_k)_\mu (P-p_k)_\nu -
\delta_{\mu\nu} (P-p_k)^2 \right) \; \times
$$
\begin{equation}
\frac{1}{2M_{H_Q}}
\langle H_Q(P)|\,\bar{Q}\gamma_\mu q(0)\,|k(\vec p_k)\rangle
\langle k(\vec p_k)|\, \bar{q}\gamma_\nu Q(0)\, |H_Q(P)\rangle
\;.
\label{B78}
\end{equation}
The factor of $2$ above corresponds to two possible directions of $\vec
p_k$ and is the unit `sphere' surface area in one space dimension.
If the smeared width is considered, the corresponding smeared spectral
density $\bar\rho(q^2)$ replaces $\rho(q^2)$ in the above equation,
since both the decay amplitudes and phase space depend on the
combination $m_Q-\sqrt{q^2}$ (for simplicity we imply that $q^2 \gg
m_q^2$).

With the smooth $\bar\rho(q^2)$, we can neglect $p_k$ compared to
$M_{H_Q}$ in Eq.\,(\ref{B78}) and have, up to $1/m_Q$ corrections,
$$
\int {\rm d}q^2\:\Gamma_k(q^2)\;=
\frac{G^2}{2 M_{H_Q}}\, m_Q^2 \bar\rho(m_Q^2) \;\times
$$
\begin{equation}
\left(v_\mu v_\nu -
\delta_{\mu\nu}\right) \:
\int \frac{{\rm d}\vec{p}_k}{2\pi\, 2E_k}
\langle H_Q(P)|\,\bar{Q}\gamma_\mu q\,|k_{\vec{p}}\rangle
\langle k_{\vec{p}}|\, \bar{q}\gamma_\nu Q \, |H_Q(P)\rangle
\;.
\label{B180}
\end{equation}
The corrections to this expression appear due to the dependence of
$\bar\rho(q^2)$ on $q^2$ near $q^2\simeq m_Q^2$ and are
power-suppressed.

If we formally summed over the final states $k$, the r.h.s.\ of the
above equation would become
$$
\frac{G^2}{2 M_{H_Q}}\, m_Q^2 \bar\rho(m_Q^2) \,
\left(v_\mu v_\nu \! - \! \delta_{\mu\nu}\right) \,
\sum_k  \int \frac{{\rm d}\vec{p}_k}{2\pi\, 2E_k}
\langle H_Q(P)|\,\bar{Q}\gamma_\mu q (0)\,|k_{\vec{p}}\rangle
\langle k_{\vec{p}}|\, \bar{q}\gamma_\nu Q (0)\, |H_Q(P)\rangle
$$
\begin{equation}
= \; \frac{G^2}{2}\, m_Q^2  \bar\rho(m_Q^2) \;
\left(v_\mu v_\nu -\delta_{\mu\nu}\right) \:
\frac{1}{2 M_{H_Q}}\,
\langle H_Q(P)|\,(\bar{Q}\gamma_\mu q)\,
(\bar{q}\gamma_\nu Q)(0)\, |H_Q(P)\rangle
\;,
\label{B182}
\end{equation}
which shows the representation of the matrix element as a sum over
intermediate states. The integration over $q^2$ replaces
${\rm d}q^2/(2\pi M_{H_Q}|\vec p_k|)$ in the partial decay width by
${\rm d}\vec p_k/(2\pi\, 2E_k)$ which is just the quantum mechanical
summation over the intermediate states.

Of course, this correspondence is not accidental for $D=2$. Indeed, the
general two-body phase space is given by
\begin{equation}
\Phi_2(M_{\!H_{\!Q}}^2,q^2\!,M_k^2) =
\int \! \frac{{\rm d}^{D} l}{(2\pi)^{\!D}}
\int \! \frac{
{\rm d}\Omega_{\vec{p}} \,
{\rm d}\mvec{p}^{\!D\!-\!1}  {\rm d}p_0}
{(2\pi)^D }
(2\pi)^2 \delta_+(l^2\!-\!q^2) \delta_+(p^2\!-\!M_k^2)
(2\pi)^{\!D}\delta^{\!D}\!(\!P\!-\!p\!-\!l)
.
\label{B184}
\end{equation}
To obtain the decay width we multiply it by the transition amplitudes
$\langle H_Q|\bar{Q}\gamma_\mu q|k\rangle$,
$\langle k|\bar{q}\gamma_\nu Q |H_Q\rangle$,
by $\frac{1}{\pi}\Im \Pi_{\mu\nu}(q)$ and by
$(2\pi)^D/(2 M_{H_Q})$. Integrating over $q^2$, we get
$$
\int {\rm d}q^2\:\Gamma_k(q^2)\; = \; \frac{G^2}{2}\,
\frac{1}{2M_{H_Q}}\,\int {\rm d}q^2\:
\Phi_2(M_{H_Q}^2,q^2,M_k^2)\;
\langle H_Q|\bar{Q}\gamma_\mu q|k_{\vec{p}}\rangle
\langle k_{\vec{p}}|\bar{q}\gamma_\nu Q |H_Q\rangle
\;=
$$
\begin{equation}
= \;G^2\,
\int \frac{{\rm d}^{D-1}\vec{p}_k}{(2\pi)^{D-1} \,2E_k}
\: \frac{1}{\pi} \Im \Pi_{\mu\nu}(P-p_k)\:
\frac{1}{2M_{H_Q}}\,\langle H_Q|\bar{Q}\gamma_\mu q|k_{\vec{p}}\rangle
\langle k_{\vec{p}}| \bar{q}\gamma_\nu Q |H_Q\rangle
\;.
\label{B186}
\end{equation}
With $\Im \Pi_{\mu\nu}(P-p_k)$ approximated by its (smeared) value at
$q=m_Q v$, this is the contribution of the given state $k$ with mass
$M_k$ (whether meson, baryon, or a multi-particle state) to the
expectation value of the local four-quark operator
$\langle
H_Q(P)|(\bar{Q}\gamma_\mu q) (\bar{q}\gamma_\nu Q) |H_Q(P)\rangle$
\cite{WA}.

Of course, in reality we do not want to extend the summation to {\it
all} states with arbitrary large energies, which would correspond to
integrating over $q^2$ down to $q^2 = -\infty$. The large-$q^2$
domain is cut at some $q^2 \lsim (m_Q-\Delta)^2 \simeq
m_Q^2-2m_Q\Delta$, with $\Delta \gsim \Lam$. This means that the heavy
quark expectation value in Eq.\,(\ref{B70}) corresponds to the
normalization point $\mu=\Delta$. We note that the normalization point
enters as the cutoff over the {\it energy} of the intermediate states, the
common case for the heavy flavor systems (see reviews \cite{rev,varenna}
and references therein).

It is worth reiterating an important point regarding the end-point
domain: the matrix elements in Eqs.\,(\ref{B70}), (\ref{B182}),
(\ref{B186})
are those in the effective theory rather than in full QCD -- even if
there were no high-momentum gluons in the latter. The representation of
the four-fermion expectation values
\begin{equation}
\langle H_Q(P)|
(\bar{Q}\gamma_\mu q)\,(\bar{q}\gamma_\nu Q) |H_Q(P)\rangle\;=\;
\sum_k\,
\int \frac{{\rm d}^{D-1}\vec{p}_k}{(2\pi)^{D-1}\,2E_k}
\langle H_Q|\bar{Q}\gamma_\mu q|k_{\vec{p}}\rangle
\langle k_{\vec{p}}|\bar{q}\gamma_\nu Q |H_Q\rangle
\label{B86}
\end{equation}
in the full theory includes not only the intermediate states without
heavy quark $Q$, Fig.\,8a, but also those containing the pair of $Q$ and
$\bar{Q}$, Fig.\,8b. The latter are not related to the
(perturbative) heavy quark
loops and appear even in the free theory -- for example, the state
$H_Q(P)\,+\, (\bar{Q}+q)$ with $\bar{Q}$ and $q$ having small
spacelike momenta $\lsim \Delta$ totalling to $\vec{p}_k$. It can be
simply $H_Q(P)+ \tilde H_Q(\vec{p}\,)$. These states appear
since the Lorentz-covariant currents $\bar{Q}\Gamma q$, together with
operators creating $Q$ contain also
operators annihilating antiquark $\bar{Q}$, which is not the case in the
nonrelativistic field theory. Such amplitudes are shaped by the time
scales $\sim 1/\Delta$ and are as nonperturbative as
transitions into the usual light hadrons. Imposing the cut on the
virtuality itself does not help to eliminate such spurious states with
``valence'' $Q$. In particular, integrating out the high-momentum modes
of quark and gluon fields, or imposing a cut on the gluon
momentum (virtuality) in the full theory -- while generating an
effective low-energy theory -- would not yield relevant operators, at
least in this context.

\thispagestyle{plain}
\begin{figure}[hhh]
 \begin{center}
 \mbox{\epsfig{file=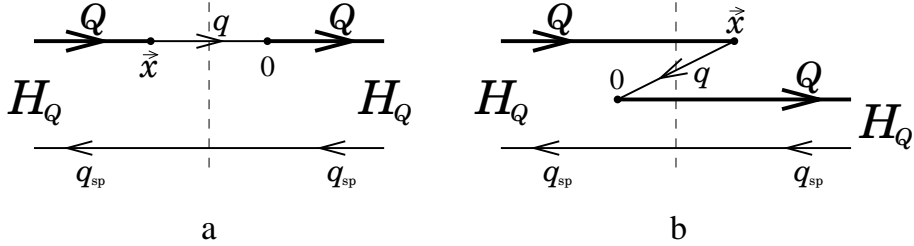,width=12cm}}
 \end{center}
 \caption{
``$s$-channel'' with $n_Q=0$ (a) and ``$u$-channel'' with $n_Q=2$ (b)
intermediate states saturating the four-quark expectation values in full
QCD. Both types are described by `soft' transition amplitudes
and contribute simultaneously to the forward ${\cal O}(G_F^2)$
transition amplitude near maximal $q^2$. Only the targeted
``$s$-channel''
states appear in the nonrelativistic effective theory. The two points
with weak currents are shown separated in space ($\vec{x}\ne 0$) for
clarity.
}
\end{figure}

The extra contributions of Fig.\,8b are counterpart of the intermediate
states in the transition operator which are nearly on-shell for the
reverse ordered product of currents
$(\bar{q}\gamma_\nu Q)(\bar{Q}\gamma_\mu q)$ in their time-ordered
product,
\begin{equation}
\hat{T}_{\mu\nu}(q)\;=\;
\int {\rm d}^D x \: {\rm e}\,^{-iqx} \;
i\,T\,\left\{\bar{q}\gamma_\nu Q(0)\, \bar{Q}\gamma_\mu q(x)
\right\}
\label{B90}
\end{equation}
at $x_0<0$. They are responsible for the $u$-channel cut of the
Lorentz-covariant forward transition amplitude $T_{\mu\nu}$ in
Eq.\,(\ref{16}). In the kinematics $q^2 \to m_Q^2$,
$q_0 \to m_Q$ the transitions
to both $s$ channel states with $n_Q=n_{\bar{Q}}=0$ and $u$ channel
states with $n_Q=n_{\bar{Q}}=1$ are equally `soft' and nonperturbative,
and cannot be disentangled in the single amplitude $T_{\mu\nu}(q)$.

On the contrary, only the necessary light intermediate states are
present when the proper nonrelativistic effective theory of heavy quarks
is considered, where $Q(x)$ do not include antiquark operators, either
creation or annihilation. The expectation values determining the
corrections to the widths must be understood only in this sense.

We conclude this Appendix by noting that the leading-$m_Q$ transition
amplitudes $H_Q \to k$ obey the stated scaling in respect to
$m_Q$ and $q^2$, {\it i.e.}, for the particular final state $k$ depend on
one combination $(m_Q^2+M_k^2-q^2)/m_Q$ having the meaning of the
rest-frame energy $E_k$:
\begin{equation}
\frac{1}{\sqrt{2m_Q}}
\langle k|\bar{q}\Gamma Q |H_Q\rangle_{q^2, m_Q}\;\simeq
\frac{1}{\sqrt{2m_{Q\prime}}}
\langle k|\bar{q}\Gamma Q' |H_{Q\prime}\rangle_{{q\prime}^2, m_Q\prime}
\label{B100}
\end{equation}
if
$$
\frac{m_Q^2+M_k^2-q^2}{m_Q}=\frac{{m_{Q\prime}}^2+M_k^2-{q\prime}^2}
{m_{Q\prime}}
\;.
$$
This functional dependence ensures, for example, that the four-quark
expectation values discussed above have a finite $m_Q$-independent
limit as $m_Q \to \infty$.

In terms of the light-cone parameters Eq.\,(\ref{B100}) says that the
formfactors, up to certain powers of $\sqrt{m_Q}$ must be functions
of $(1\!-\!\omega) m_Q$, see relation
(\ref{B4a}). The explicit expressions (\ref{B5})--(\ref{B9}) do exhibit
this property,
although in slightly different ways in different kinematic domains.
To locate it, we once again can be guided by the duality between the
various pieces in the hadronic expressions based on the 't~Hooft
equation, and the contributions coming from the corresponding Feynman
diagrams. This duality exists in the physical gauge and is transparent
when the light-cone formalism is used. Additionally, the scaling
behavior of various perturbative effects can be anticipated beforehand
if general OPE facts are considered. For example, the high infrared
stability of the inclusive widths, together with the fact that no gluon
degrees of freedom (and, therefore, no bremsstrahlung) exist in $D=2$,
leads to the finiteness of the vertex corrections and allows to
estimate their magnitude.

It has been mentioned already that Eq.\,(\ref{B100})
holds at large energy $E_k \gg
\beta$: then the amplitudes are given only by the simple overlaps ${\cal
C}_{k}$ and ${\cal D}_{k}$. The integrals in Eqs.\,(\ref{153}) and
(\ref{154}) manifestly depend only on $(1\!-\!\omega) m_Q$ if
$\Psi_{H_Q}(t)$ are nonzero only for $t\lsim \beta$.

A different situation takes place at maximal $q^2$, for example, for
light final state quarks. The triple-overlap terms proportional to
$F_{lk}$ are then of the same order. However, in this kinematics
$F_{lk}$ show the same dependence on $(1\!-\!\omega) m_Q$. Indeed, in
this case $\omega \to 1$, and they take the form
\beq
F_{lk}(\omega)\; \simeq\; -\,m_Q^2(1\!-\!\omega)\;
\int_0^{\infty} du\,\int_0^1 {\rm d}y\,
\frac{\Psi_l^{B_c}(u)\varphi_k(y)}{[m_Q(1\!-\!\omega)y+u]^2}
\Psi_{H_Q}(\,m_Q(1\!-\!\omega)(1\!-\!y)\,)
\;.
\label{B102}
\eeq
The position of the excited states in the $t$ channel is also driven by
the nonrelativistic expression $\mu_l \simeq m_Q+ \tilde \epsilon_l$
with $m_Q$-independent $\tilde \epsilon_l$. Therefore, the pole factors
in front of $F_{lk}$ are simply proportional to $1/m_Q$:
\beq
\frac{q^2,\;\,\mu_l^2}{q^2-\mu_l^2}\;\simeq \;
\frac{m_Q}{-2(\Delta + \tilde\epsilon_l)} \qquad \mbox{for }\,
q^2=(m_Q-\Delta)^2 \;\; \mbox{with }\,\Delta \ll m_Q
\;.
\label{B104}
\eeq
On the other hand, $\Delta$ differs from $(1\!-\!\omega) m_Q$ by just
a ($M_k$-dependent) constant,
$$
E_k\:\simeq \:M_{H_Q}-m_Q\,+\, \Delta \:=\:
\epsilon_0+ \Delta\,,\qquad \qquad
E_k+\sqrt{E_k^2-M_k^2} = (1\!-\!\omega) m_Q\;.
$$
Thus, together with the factors $c_l\propto m_Q^{-1/2}$ this term
possesses the required dependence on $(1\!-\!\omega) m_Q$ and
overall power scaling in $m_Q$.

This reasoning may seem inapplicable when both $Q$ and $q$ are heavy
quarks, in particular, at small $m_Q-m_q$. The $t$-channel $B_c^{(l)}$
wavefunctions are then nonrelativistic in respect to both quarks, and
must be treated accordingly. This, however, is
a transition between two heavy quarks, and even the classic case of the small
velocity \cite{volshif}. As mentioned above, in $D=2$ the amplitudes are
finite even at $\vec{v}\ne 0$ and the perturbative corrections are
inversely proportional to the heavy quark mass (but not to powers of
$1/q^2$ which blows up at the safe zero recoil point).
And indeed, it is easy to find that $F_{lk}$ in Eq.\,(\ref{152}) are
universally suppressed by the mass scale $E_k\sim (1\!-\!\omega) m_Q$
whether $m_q$ is small or large.

Therefore, the functional relation Eq.\,(\ref{B100}) holds in the
't~Hooft model for any decay kinematics.

\subsection{The IW function and the heavy quark distribution functions}
\renewcommand{\theequation}{A3.\arabic{equation}}
\setcounter{equation}{0}

The IW function $\xi$ is most simply defined as the flavor-diagonal vector
formfactor
$$
\langle H_Q(p')|\,\bar{Q}\gamma_\mu Q\,|H_Q(p)\rangle =
\xi\left(\frac{(pp')}{M_{H_Q}^2}\right)\, (p+p')_\mu
$$
in the large-$m_Q$ limit. The expression for it quoted below can be
obtained using
the $1/m_Q$ expansion of the amplitudes in the previous section, if we
employ the nonrelativistic expansion of the final state as well.
Alternatively, we can use directly the simple universal expressions
Eq.\,(\ref{105}) at $q^2=0$ relying on the scaling relations
(\ref{134})--(\ref{136}) by adjusting the final-state quark mass:
\begin{equation}
M_{H_{Q'}}\;=\; z M_{H_Q} \qquad \mbox{with} \qquad
w\equiv (vv') = \frac{1+z^2}{2z}\,, \qquad z=w\pm \sqrt{w^2-1}
\;.
\label{C4}
\end{equation}
The perturbative corrections are finite and vanish at $m_Q \to \infty$ in
two dimensions. In this way we obtain
$$
\xi(w)\;=\; \frac{2}{1+w+\sqrt{w^2-1}}\, \int_0^\infty {\rm d}t\:
\Psi_{H_Q}\left(t\right) \Psi_{H_Q}\left([w - \sqrt{w^2-1}]t\right)\;=
$$
\begin{equation}
=\;\frac{2\sqrt{z}}{1+z} \, \int_0^\infty {\rm d}t\:
\Psi_{H_Q}\left(\frac{t}{\sqrt{z}}\right)
\Psi_{H_Q}\left(\sqrt{z} t \right)
\;.
\label{C6}
\end{equation}
The last expression was used to make representation more symmetric. Here
$\Psi_{H_Q}(t)$ is the solution of the static 't~Hooft equation
\beq
\epsilon \:\Psi(t) \;=\;
\left(\frac{m_\sp^2-\beta^2}{2t}\, + \,\frac{t}{2}
\right) \, \Psi(t)\;-\;
\frac{\beta^2}{2}\, \int_{0}^{\infty} {\rm d}s\:
\Psi(s)\,\frac{1}{(t-s)^2}
\;\;.
\label{C8}
\eeq
Expression (\ref{C6}) is easily generalized for the case of inelastic
transitions:
\begin{equation}
\xi^{kn}_{\rm inel}(w)\;=\;
\frac{2\sqrt{z}}{1+z} \, \int_0^\infty {\rm d}t\;
\Psi^{(k)}_{H_Q}\left(\frac{t}{\sqrt{z}}\right)
\Psi^{(n)}_{H_Q}\left(\sqrt{z} t \right)
\;.
\label{C9}
\end{equation}
This agrees with the expression obtained in Ref.\,\cite{burkswan}.
We found a direct proof (it will be presented elsewhere) that the
right-hand side of Eq.\,(\ref{C9}) does not change (up to a sign) if
$z\to 1/z$. This ensures the same result for both solutions for $z$ in
terms of $w$, the last of Eqs.\,(\ref{C4}).

It is interesting to note that the absolute maximum for the (elastic) IW
function which would correspond to the structureless point-like heavy
flavor hadron, is given by just the factor~\footnote{This
function is not analytic and rather has a branch point near $w=-1$ (or
$q^2=4m_Q^2$). This is just a reflection of the Fermi statistics of heavy
quarks confined in the bosonic state. This means that it is impossible
to construct a heavy pointlike meson from fermionic constituents, with
the radius much smaller than the mass. For light constituents this is
possible as exemplified by the chiral pion.}
\begin{equation}
\xi_{\rm max}(w)\;=\; \sqrt{\frac{2}{1+w}}=
\frac{2\sqrt{z}}{1+z}\;.
\label{C12}
\end{equation}
It would be saturated by the light cone wavefunction $\Psi(t) \sim
c/\sqrt{t}$. Indeed, one has
\begin{equation}
\xi(z)\;=\; \left(\frac{2\sqrt{z}}{1+z}\right)^{2+2\alpha}
\qquad \mbox{for} \qquad \Psi(t)=t^\alpha {\rm e}\,^{-\mu t}
\;.
\label{C14}
\end{equation}
Of course, such a situation is not realized for any actual hadronic
state in the 't~Hooft model. Constructing such a hard-core wavefunction
requires a coherent superposition of all the excited eigenstates.
\vspace*{.2cm}

Studying the decay distribution for the weak transitions into light
quarks, $m_c^2 \ll m_Q \Lam$ allows one to determine the light-cone
distribution function. In particular, it emerges directly in the decays
at $q^2=0$ as the differential decay probability vs.\ $q_0$ or
$M^2_{\rm hadr}$
\cite{motion}:
\begin{equation}
\frac{1}{\Gamma_{\rm sl} (0)}
\frac{{\rm d}\Gamma_{\rm sl} (0)}{{\rm d}M^2_{\rm hadr}}\;
\stackrel{m_Q\to\infty}{=}\;
m_Q (M_{H_Q}-m_Q)\,
F\left(1-\frac{M^2_{\rm hadr}}{m_Q(M_{H_Q}-m_Q)}
\right)\;,
\label{C20}
\end{equation}
$$
q_0=\frac{M_{H_Q}^2-M_{\rm hadr}^2}{2M_{H_Q}}\;\mbox{ at }
q^2=0\;,
$$
where $\Gamma_{\rm sl} (0)$ as a function of $M^2_{\rm hadr}$ is
$$
\sum_{M_k^2 < M^2_{\rm hadr}} \Gamma_{\rm sl}^{(k)} (0)\;.
$$
Using the leading term in the $1/m_Q$ expansion of the decay amplitudes,
one can check that $F(t)$ coincides with $\Psi_{H_Q}^2(t)$. To
obtain the distribution over $M_k^2$, it is again convenient to use the
explicit cutoff-dependence of the ultraviolet-regularized Green function
$G(x,y;\Lambda)$, as it was done
in Sect.\,3.  Earlier discussion for the 't~Hooft
model  can be found in Refs.\,\cite{einhorn,burk}.

The small-$x$ behavior of the heavy-quark
distribution function for decays into light quarks has, in general, a
non-integer power depending on the spectator mass,
\begin{equation}
F(x)\; \propto\; x^{2\gamma_{\rm sp}}\:, \qquad
\frac{\pi \gamma_{\rm sp}}{{\tan}{\pi \gamma_{\rm sp}}} =
-\frac{m_{\rm sp}^2-\beta^2}{\beta^2}
\;.
\label{C22}
\end{equation}
Just such a behavior was conjectured for the QCD light-cone distribution
function in the model suggested in \cite{bsg}. In $D=4$, however, the
distribution function itself and the exponent $2\gamma$, in particular
depend on the renormalization point.

\subsection{Perturbative corrections and IR regularization}
\renewcommand{\theequation}{A4.\arabic{equation}}
\setcounter{equation}{0}

The perturbative corrections to the weak decay vertex explicitly depend
not only on the gauge but also on the employed infrared regularization,
even at $q^2=0$ (see, e.g.\ Ref.\,\cite{D2}, Sect.\,III\,\,B). This
subtlety is aggravated by severe infrared divergences in the individual
diagrams in $D=2$, which appear when unphysical degrees of freedom are
introduced (say, in the covariant gauges). While OPE -- applicable for
inclusive widths -- ensures that the widths are rather insensitive to
the infrared contributions, in practical computations it requires
carrying out one and the same regularization procedure consistently
through all stages of computations. A particular infrared
regularization, on the one hand, is built into the 't~Hooft equation in
the form of the principal value  prescription of the Coulomb exchange
integral and the self-energy terms. On the other hand, the same infrared
regularization was employed in establishing the short-distance
non-renormalization theorem for the weak vertex.

Therefore, it is instructive to obtain the same perturbative correction
$\beta^2/(2m_Q^2)$ to the decay width starting with the usual Feynman
graphs in the covariant gauge routinely used in four dimensions. To
this end we independently calculated the one-loop perturbative
corrections to the quark decay width $Q \to q+\phi$ at $m_\phi^2=0$,
with the decay interaction
\begin{equation}
{\cal L}_{\rm weak} \;=\; -\frac{G}{2\pi} \:\bar{q} Q\, \phi \;+\;
{\rm H.c.}
\;.
\label{D5}
\end{equation}
In $D=2$ this is a full analogue of the four-fermion (semileptonic)
decay with massless leptons for vector-like weak currents.

Since all the individual diagrams are too infrared divergent, the
computations are simple only in dimensional regularization. All
corrections are nontrivial, including virtual vertex corrections, mass
and wavefunction renormalization and the ``real'' gluon emission width.
The sum of all contributions takes the following form (at $m_q=0$) in
arbitrary dimension:
$$
\frac{\delta \Gamma^{\rm 1-loop}}{\Gamma^{\rm tree}} \;=\;
C_F \frac{g_s^2}{(4\pi)^{\frac{D}{2}}}\; m_Q^{D-4}\:
\left\{
\Gamma\left(2-\frac{D}{2}\right)
\left[
4\frac{D-3}{D-4}+\frac{(D-1)(D-4)}{D-3}
\right] \;+
\right.
$$
\begin{equation}
\left.
+\;
3\,\frac{\Gamma^2\left(D/2\right)}{\Gamma\left(\frac{3}{2}D-2\right)}
\:\frac{3D^4-28D^3+109D^2-208D+160}{(D-1)(D-3)(D-4)^2}
\right\}
\;.
\label{D7}
\end{equation}
It is easy to see that the poles at $D=2$ and $D=3$ all cancel out (at
$D=3$ heavy quark masses acquire infrared logarithmic divergence as well).
The remaining pole at $D=4$ is given by the ultraviolet renormalization
of the Yukawa coupling.

At $D=2$ the above expression reduces to
\begin{equation}
\delta \Gamma^{\rm 1-loop}\;=\;
C_F \frac{g_s^2}{4\pi m_Q^2} \:\Gamma^{\rm tree}
\;,
\label{D9}
\end{equation}
which coincides exactly with the result obtained in the light-front
formalism and by the summation of the exclusive hadronic widths in the
't~Hooft model.

We note that here one has an instructive example: the
perturbative calculation of the decay width is IR safe by itself, and
the first-order correction is meant to describe the actual decay width
with the $1/m_Q^2$ accuracy. Higher orders of the perturbative expansion
yield $1/m_Q^4$ and smaller terms, and the uncertainties in summing
the perturbative series are exponentially small in $m_Q$.
However, this is not the complete
answer already to the order $1/m_Q^2$. This illustrates incompleteness
of the purely perturbative expansion, even in that the actual
nonperturbative effects are not signalled here by generic intrinsic
divergences of the perturbative series.

\end{document}